# Fast BSS Transition for Multi-PSK Community Wi-Fi in MDU Networks: Fragmented RADIUS Key Delivery, Server-Derived PMK-R0/PMK-R1 Distribution through a Cloud Management API, and Lease-Driven Key Revocation

**Azif Abdul Salam Kolisam beevi**[1,*], **Anil Kumar Reddy**[2]

[1]Director of Enterprise Solutions Engineering, San Jose, CA, USA
[2]Principal Engineer, Bangalore, Karnataka, India
[*]Corresponding author: azif.abdulsalam@gmail.com
Co-author email: anilgeereddy@gmail.com

## Abstract

Multi-dwelling-unit (MDU) properties increasingly deploy a single **community-wide SSID** on which every unit or resident has its own pre-shared key, and each key is mapped to its own VLAN so that every unit is a separate network. Vendors call this capability by different names (ePSK, iPSK, MPSK, DPSK, PPSK); we refer to it generically as **multi-PSK (mPSK)**. When the keys are held in an external RADIUS server or a cloud controller, the access point (AP) cannot identify a client's key by itself and must ask the key server. Enabling IEEE 802.11r Fast BSS Transition (FT) with WPA2-PSK (FT-PSK) on such an SSID, as large properties need for roaming voice and video, breaks this exchange in three ways. First, the server must identify the key from an FT handshake whose MIC depends on the FT key hierarchy. Second, the RADIUS messages no longer fit: the EAPOL-Key message 2 grows to 239–290 octets and the key material the AP needs grows to 258–354 encrypted octets, both beyond the 247-octet value capacity of a RADIUS vendor-specific attribute (VSA). Third, the AP a device roams to needs a PMK-R1 for that device, and all of the key material must disappear when the resident leaves. We present **FT-mPSK**. The key server identifies the PSK, derives PMK-R0 and PMK-R1, and returns what the serving AP needs to complete the handshake in a **two-fragment Key-Bundle VSA** that the AP reassembles and authenticates before decryption. It then calls **one POST API on the cloud management system**, which pushes PMK-R0 to the primary AP. When the device roams, the target AP obtains its PMK-R1 in one of two ways before answering the FT Authentication Request: from the primary AP, which it finds from the R0KH-ID in the request, or from the key server. When the **Property Management System (PMS)** reports a move-out or lease end, the key server issues the same POST with a delete operation keyed by the resident's **mPSK username**, and every AP purges that username's PSK, PMK, PMK-R0 and PMK-R1. We specify both the external-RADIUS and cloud-controller deployment methods, the attribute dictionary and fragment format, the API, the algorithms, size budgets, complexity and security properties, and the laboratory validation plan.

## 1. Introduction

Apartment communities, student housing, senior living and extended-stay hotels have moved from per-unit Wi-Fi routers to a managed, property-wide wireless network. The prevailing design is a **single community-wide SSID** that residents can use anywhere on the property, in their unit, the gym, the pool or the leasing office. Security and privacy come from giving every unit, or every resident, a **unique pre-shared key** and mapping each key to its **own VLAN**, so that a resident's devices see only one another and the Internet. Vendors implement this capability under different names, such as ePSK, iPSK, MPSK, DPSK and PPSK [1]. In this paper we use the generic term **multi-PSK (mPSK)**.

In large properties the keys are not configured on the APs. They are held in an **external RADIUS server** [2], often integrated with the property's leasing platform, or in a **cloud controller** that manages the APs. The AP cannot tell from the association which of thousands of keys a client is using. It therefore forwards the client's handshake to the key server, which finds the key whose PMK reproduces the MIC of EAPOL-Key message 2 (M2) and returns the key and the unit's VLAN.

Large properties also need **IEEE 802.11r Fast BSS Transition (FT)** [3], [4]: residents walk through corridors and common areas on voice and video calls, and a full four-way handshake with a round trip to an external key server at every AP change is too slow. With WPA2-PSK, FT is negotiated as **FT-PSK (AKM 4)**. Enabling FT-PSK on an mPSK SSID with an external key store raises four problems that existing mPSK designs do not address:

1) **Key identification changes.** In an FT initial mobility-domain association, the M2 MIC is computed from a key hierarchy (PMK → PMK-R0 → PMK-R1 → PTK) that depends on the mobility domain and the key-holder identities. The server must evaluate this hierarchy for each candidate key.
2) **The RADIUS messages no longer fit.** The FT M2 carries the Mobility Domain element (MDE) and the Fast BSS Transition element (FTE) and grows to 239–290 octets. The key material the AP needs to complete the handshake (PSK, PMK, PMK-R1 and the key names) grows to 258–354 octets once encrypted. Both exceed the 247-octet value capacity of a RADIUS vendor-specific attribute (VSA), so **the key server cannot return the PSK and its FT keys in a single attribute; they must be fragmented**.
3) **The target AP needs keys.** For a client to roam with one FT exchange, the target AP must have the PMK-R1 derived for it before it answers the client's FT Authentication Request. With keys in an external store, the target AP needs a fast and secure way to obtain it.
4) **Keys must disappear when residents leave.** When a lease ends, the resident's PSK, and every PMK, PMK-R0 and PMK-R1 derived from it on every AP, must be removed promptly.

This paper presents **FT-mPSK**, an end-to-end design that solves these problems for both the external-RADIUS and the cloud-controller deployment methods. Its main elements are:

1) **Fragmented RADIUS attributes.** A two-fragment VSA format with a 4-octet fragment header carries the FT M2 in the Access-Request and the encrypted **Key-Bundle** in the Access-Accept. The AP reassembles and authenticates the fragments before decryption.
2) **Server-derived key hierarchy.** After identifying the key, the key server derives **PMK-R0** and the serving AP's PMK-R1. PMK-R1 travels in the Key-Bundle; PMK-R0 and the resident's binding are delivered to the primary AP through a single **POST /key-sync API** on the cloud management system (CMS), sealed end-to-end to that AP.
3) **On-demand PMK-R1 delivery at roam time, in two approaches.** In **Approach 1**, the target AP reads the R0KH-ID of the primary AP from the client's FT Authentication Request and pulls PMK-R1 from it. In **Approach 2**, the target AP requests PMK-R1 from the RADIUS server or cloud key service, which re-derives it for that AP. In both cases the target AP then sends the FT Authentication Response.
4) **Lease-driven revocation.** The PMS notifies the key server of move-outs, lease ends and transfers, and the key server issues a delete through the same POST API, keyed by the resident's **mPSK username** (for example, the ePSK username). Every AP then purges all PSK, PMK, PMK-R0 and PMK-R1 associated with that username and disconnects the resident's devices.
5) **Per-PSK VLAN micro-segmentation** is preserved through FT roams, because the VLAN travels with PMK-R1 to every AP the client roams to.

Section 2 reviews mPSK, per-PSK VLANs, key identification, FT and the deployment models. Section 3 states the problem. Section 4 presents the architecture, attribute dictionary, fragment format and API, and Section 5 the algorithms. Sections 6–8 analyse sizes, complexity and security. Section 9 describes the evaluation, Section 10 discusses limitations, and Section 11 concludes.

## 2. Background

### *2.1 Multi-PSK: One SSID, Many Keys*

In conventional WPA2-Personal all clients of an SSID share one passphrase. An mPSK SSID accepts many passphrases at once. Each passphrase belongs to a unit, a resident or a device, and the network learns which one a client used during its four-way handshake. The concept is the same across vendors, but the names differ. Table 1 lists common terms; throughout this paper **mPSK** stands for all of them. Each key is also provisioned with a **username**, for example the ePSK username, that identifies the unit or resident account. The PMS, the key server and the APs refer to a key, and to everything derived from it, by this username.

**Table 1.** Vendor terminology for multi-PSK (mPSK) [1]

| Term | Expansion | Example vendors |
|---|---|---|
| iPSK | Identity PSK | Cisco, Cisco Meraki |
| MPSK | Multiple PSK / Multi-PSK | HPE Aruba, Fortinet |
| DPSK | Dynamic PSK | Ruckus (CommScope) |
| PPSK | Private PSK | Extreme Networks (incl. Aerohive), Juniper Mist, Ubiquiti UniFi |
| ePSK | Enterprise PSK | Cambium Networks |
| mPSK (this paper) | multi-PSK, generic | All of the above |

### *2.2 Per-PSK VLAN Micro-Segmentation*

The property value of mPSK comes from pairing each key with a VLAN. When the key server identifies PSK_k, it returns VLAN_k, in RADIUS as Tunnel-Type = VLAN, Tunnel-Medium-Type = IEEE-802 and Tunnel-Private-Group-ID = VLAN_k [5], [6], or in the cloud controller's response. The AP tags all of the client's traffic with VLAN_k [7]. All devices of one unit (phones, laptops, smart TVs, speakers, IoT) share the unit's PSK and therefore its VLAN, which forms a **personal area network**: devices in the unit can discover one another, but not the devices of other units. The gateway or firewall isolates the VLANs and permits only Internet access between them. Figure 1 illustrates the model.

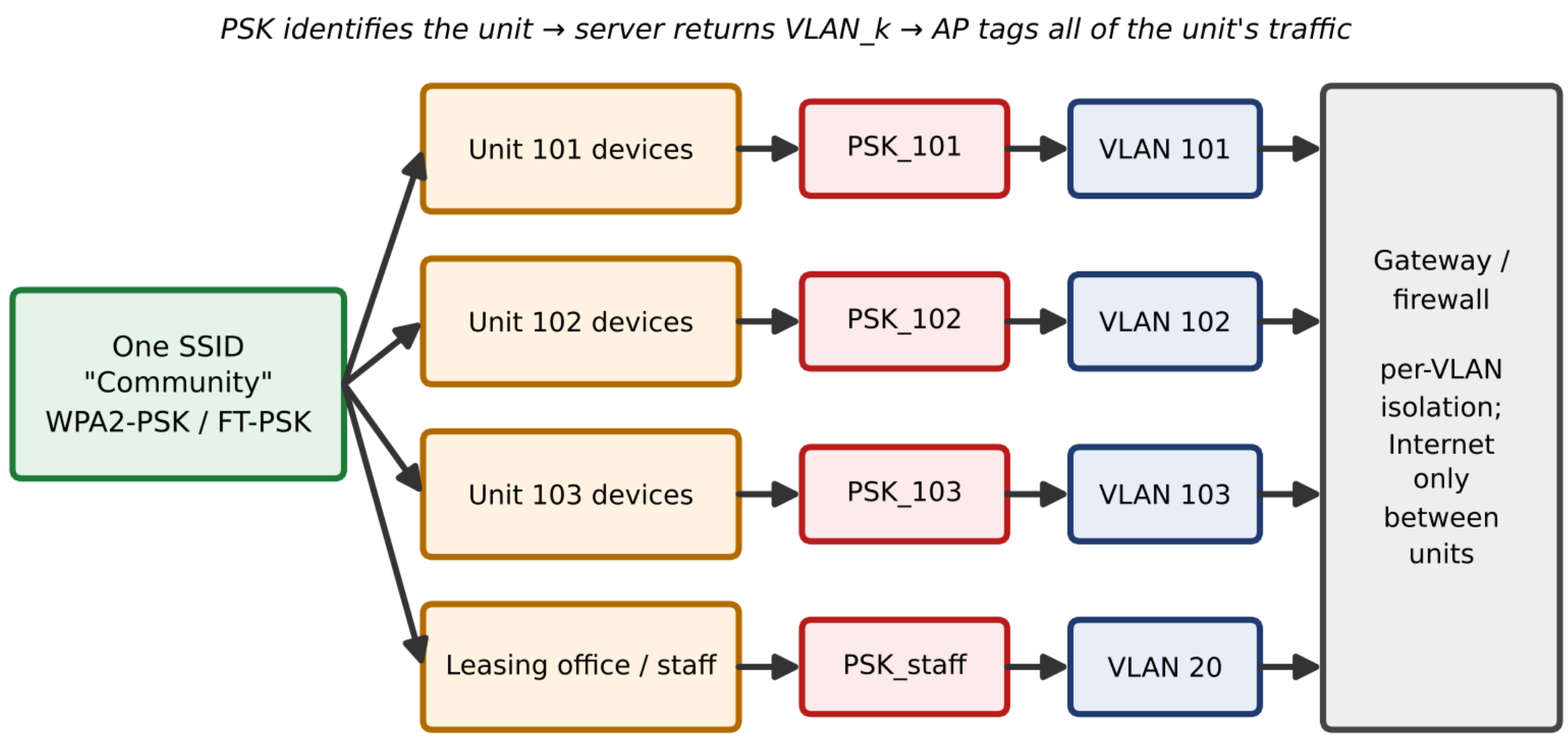

**Figure 1.** Per-PSK VLAN micro-segmentation on one community-wide SSID: the PSK identifies the unit, the key server returns that unit's VLAN, and the AP tags all of the unit's traffic with it.

Three practical points matter for the design. First, the VLAN must be known at every AP the client may roam to, so it travels with the key material. Second, the 12-bit VLAN identifier limits a layer-2 domain to 4,094 VLANs; larger properties split the SSID into several mobility domains or map units to tunnels or group policies instead of VLANs, which does not change the key handling described here. Third, all clients of a BSS share one group temporal key (GTK) regardless of VLAN, so APs should not transmit group-addressed frames of one unit's VLAN to other units' clients (for example, by multicast-to-unicast conversion and proxy ARP/ND), as discussed in Section 8.

## 2.3 Identifying a Resident's Key

In WPA2-Personal the PMK is derived from the passphrase and SSID [8], and the four-way handshake binds it to both nonces and addresses:

$$PMK = PBKDF2\text{-}HMAC\text{-}SHA1(PSK,\ SSID,\ 4096,\ 256) \tag{1}$$

$$PTK = PRF\text{-}384(PMK,\ \text{"Pairwise key expansion"},\ min(AA,SPA) \,\|\, max(AA,SPA) \,\|\, min(ANonce,SNonce) \,\|\, max(ANonce,SNonce)) \tag{2}$$

$$MIC = HMAC\text{-}SHA1\text{-}128(KCK,\ EAPOL\text{-}Key\ M2\ with\ MIC\ field\ zeroed),\ KCK = PTK[0..127] \tag{3}$$

Two identification models are in use. In **MAC-based lookup**, the AP asks the server for the key registered to the client's MAC address before the handshake. This requires every device to be registered in advance, which is impractical for the many devices residents bring. In **handshake-based identification**, the AP holds message 3 (M3), sends ANonce and the raw M2 to the server, and the server evaluates Eqs. (2)–(3) for each candidate PMK until the MIC matches. Every input except the PMK is visible in M1 and M2, and the PMK of each key is pre-computed per SSID. Handshake-based identification needs no device registration and is the model used here; a known MAC can still serve as a hint that tells the server which key to try first.

## 2.4 IEEE 802.11r Fast BSS Transition with WPA2-PSK

FT [3], [4] lets a client move between the APs of a *mobility domain*, identified by the MDID in the MDE, with a single FT authentication exchange over the air or over the distribution system (DS), instead of a full authentication and four-way handshake. With WPA2-PSK the FT AKM is **FT-PSK** (AKM 4,

SHA-256; AKM 19 is the SHA-384 variant). The first association in a mobility domain is the **FT initial mobility-domain association**. The client associates advertising AKM 4 and the MDE; the AP answers with the MDE and an FTE that carries its R0KH-ID and R1KH-ID; and the four-way handshake follows. M2 carries the RSNE, the MDE and the FTE, and M3 carries the GTK, RSNE, MDE and FTE. On a later roam, the client's FT Authentication Request carries PMKR0Name in its RSNE and the primary AP's R0KH-ID in its FTE, which tells the target AP where the client's key hierarchy is rooted. The AP that performs this association is the R0 key holder (R0KH) for the client; we call it the client's **primary AP**. Every AP of the mobility domain is an R1 key holder (R1KH).

### 2.5 FT Key Hierarchy and the FT M2 MIC

For FT-PSK the root key XXKey is the PSK-derived PMK of Eq. (1). The R0KH derives:

$$PMK\text{-}R0 \parallel PMK\text{-}R0Name\text{-}Salt = KDF(XXKey, \text{"FT-R0"}, SSIDlen \parallel SSID \parallel MDID \parallel R0KHlen \parallel R0KH\text{-}ID \parallel S0KH\text{-}ID) \quad (4)$$

$$PMKR0Name = Truncate\text{-}128(SHA\text{-}256(\text{"FT-R0N"} \parallel PMK\text{-}R0Name\text{-}Salt)) \quad (5)$$

and, for every R1KH, identified by its R1KH-ID:

$$PMK\text{-}R1 = KDF(PMK\text{-}R0, \text{"FT-R1"}, R1KH\text{-}ID \parallel S1KH\text{-}ID) \quad (6)$$

$$PMKR1Name = Truncate\text{-}128(SHA\text{-}256(\text{"FT-R1N"} \parallel PMKR0Name \parallel R1KH\text{-}ID \parallel S1KH\text{-}ID)) \quad (7)$$

$$PTK = KDF(PMK\text{-}R1, \text{"FT-PTK"}, SNonce \parallel ANonce \parallel BSSID \parallel STA\text{-}ADDR) \quad (8)$$

$$MIC = AES\text{-}128\text{-}CMAC(KCK, EAPOL\text{-}Key\ M2\ with\ MIC\ field\ zeroed) \quad (9)$$

Here S0KH-ID = S1KH-ID = the client MAC, R0KH-ID is 1–48 octets and R1KH-ID is 6 octets; the MIC of Eq. (9) uses AES-CMAC [9]. One property matters for this design: Eqs. (4)–(8) are **deterministic**. Anyone holding the PMK and the public identifiers obtains the same keys and names, so a key server can derive the hierarchy on behalf of the R0KH, and can later re-derive a PMK-R1 for any AP, without any change visible to the client.

### 2.6 Deployment Models and the RADIUS Attribute Limit

mPSK keys are held in one of two places. In the **external RADIUS method**, the AP is a RADIUS client and the key server is a RADIUS server, on premises or hosted. In the **cloud controller method**, the keys live in the cloud management system that already configures and monitors the APs, and the AP queries it over its persistent, mutually authenticated management session. Many deployments combine the two, for example a cloud-hosted RADIUS service. In both cases the cloud management system exposes a northbound HTTP API [10] and can deliver data to every AP.

A RADIUS attribute is encoded as Type (1 octet), Length (1 octet) and Value, and Length covers the whole attribute, so an attribute is at most 255 octets [2]. A Vendor-Specific Attribute (Type 26) also carries a 4-octet Vendor-Id and, in the recommended format, a 1-octet Vendor-Type and a 1-octet Vendor-Length, so the usable value capacity per VSA is:

$$V_max = 255 - 2 - 4 - 2 = 247 \text{ octets} \quad (10)$$

RFC 6929 [11] defines Long Extended attribute types for longer values, but many AP RADIUS clients and servers deployed in MDU networks support only the classic VSA format. Attributes of the same type are kept in order end to end, and proxies must not reorder them [2], [12], which makes application-level fragmentation across consecutive VSAs reliable. Keys in an Access-Accept are protected with the salt-based MD5 key hiding of RFC 2548 [13] or by a secure transport such as RadSec or DTLS [14], [15].

### 2.7 Property Management Systems

A Property Management System (PMS) is the system of record for units, residents and leases: move-in, renewal, unit transfer, move-out and lease end. In mPSK deployments the PMS usually triggers key creation in the key server when a lease starts. It is equally the natural trigger for key removal, which is the part existing deployments handle least well.

## 3. Problem Statement

### 3.1 FT Changes Key Identification

Without FT, the server identifies PSK_k from M2 with Eqs. (2)–(3) and returns PSK_k (or PMK_k) and VLAN_k. With FT-PSK, the MIC of M2 is computed from a PTK derived from PMK-R1 (Eqs. 8–9), which depends on the MDID, the R0KH-ID and R1KH-ID of the serving AP, and the client MAC. The server needs these values. It finds them in the MDE and FTE of M2, and the AP confirms them in a separate attribute so that a client cannot steer the server to another AP's identity. For each candidate key the server must then evaluate the FT hierarchy instead of the plain PRF.

Once the key is found, the serving AP needs more than the PSK to send M3: it needs **PMK-R1 for its own R1KH-ID** and the key names, because the PTK of M3 derives from PMK-R1. The AP could derive them itself from the PMK, but then the key names would be generated outside the key server, which is where they must be recorded for roaming and revocation.

### 3.2 The RADIUS Messages No Longer Fit

Table 2 compares the RADIUS payloads of plain WPA2-PSK and FT-PSK against the 247-octet VSA capacity (detailed budgets in Section 6). With FT the raw M2 in the Access-Request and the encrypted Key-Bundle in the Access-Accept both exceed one VSA in typical configurations, and always in the worst case. **The PSK and its FT key material cannot be returned in a single attribute; both directions need fragmentation.**

**Table 2.** RADIUS payloads versus VSA capacity (octets)

| Payload | WPA2-PSK (AKM 2) | FT-PSK (AKM 4) | FT-PSK SHA-384 (AKM 19) |
|---|---|---|---|
| Raw EAPOL-Key M2 (Access-Request) | 121–125 (1 VSA) | 239–290 (2 fragments typical) | 239–290 (2 fragments typical) |
| Encrypted Key-Bundle (Access-Accept), worst case | 194 (1 VSA) | 322 (2 fragments) | 354 (2 fragments) |
| Encrypted Key-Bundle, 63-char PSK, 6-octet R0KH-ID, 32-char identity | 162 (1 VSA) | 258 (2 fragments) | 290 (2 fragments) |

### 3.3 Distributing Keys across the Mobility Domain

A roam to AP-2 completes with one FT exchange only if AP-2 has the PMK-R1 derived for its R1KH-ID by the time it answers the FT Authentication Request. IEEE 802.11r lets the target AP pull PMK-R1 from the R0KH named in that request. With an external key store two further questions arise: the R0KH must actually hold PMK-R0, which the key server derives, and there must be a path that works when the primary AP is unreachable. Pushing PMK-R1 proactively to every AP for every client would answer both, but it places keys on APs the client may never visit and multiplies the key material that must later be revoked.

### *3.4 Removing Keys When a Lease Ends*

When a resident moves out, the PSK must stop working and every copy of key material derived from it must be removed: the binding, PMK and PMK-R0 at the primary AP, and a PMK-R1 at every AP the client has roamed to. Disabling the PSK in the key server prevents new associations, but FT roams are served from AP-resident keys without contacting the server, so already-connected devices keep working. RADIUS dynamic authorisation [16] ends one session at one AP and does not delete replicated keys. A binding lifetime alone is not sufficient either, because early move-outs and transfers take effect at a time set by the PMS.

### *3.5 Alternatives Considered*

- **MAC-based pre-association lookup.** The server returns the PSK for a registered MAC before the handshake, and the AP derives the FT hierarchy itself. It requires every device to be registered in advance, generates key names outside the key server and provides no revocation channel.
- **AP-side derivation from a returned PMK.** The server returns only PSK and PMK (one VSA), and the AP derives PMK-R0 and PMK-R1 and pushes PMK-R1 to its peers. The request-side M2 still needs fragmentation, and distribution and revocation still depend on the primary AP. Deriving and distributing the hierarchy for every client also places unnecessary processing and messaging load on the AP, which may affect client connections.
- **RFC 6929 Long Extended attributes** [11]. The standards-track encoding for long values. It solves the size limit where both ends support it, but not key delivery to roaming targets or revocation. The fragment header defined here maps one-to-one onto it.
- **Proactive push of PMK-R1 to every AP.** The key server derives a PMK-R1 for every AP of the mobility domain and pushes them all at identification time. Roams need no fetch, but n keys are created and stored per client, most of them on APs the client never visits, and every one of them must later be revoked.
- **Proposed (FT-mPSK).** The server identifies the key from the FT M2 (fragmented) and returns the serving AP's keys in a two-fragment Key-Bundle. PMK-R0 reaches the primary AP through one POST API on the CMS. Target APs obtain PMK-R1 on demand from the primary AP (Approach 1) or the key server (Approach 2), and the same POST API deletes all keys of an mPSK username.

**Table 3.** Comparison of approaches for FT-PSK on an mPSK SSID with an external key store

| Approach | Unregistered devices | Fits RADIUS limits | Central key naming | PMK-R1 at target AP | Revocation by username |
|---|---|---|---|---|---|
| MAC-based lookup | No | Yes | No | AP-derived | No |
| AP-side derivation | Yes | No (M2) | No | Pushed by AP (AP load) | No |
| RFC 6929 Long Extended | Yes | Yes (if supported) | Yes | Not addressed | No |
| Proactive push to every AP | Yes | Yes | Yes | Pushed to all n APs | Yes |
| **FT-mPSK (proposed)** | **Yes** | **Yes (fragments)** | **Yes** | **On demand (Approach 1 or 2)** | **Yes (POST delete)** |

## 4. Proposed Architecture

Figure 2 shows the FT-mPSK architecture. A single community-wide SSID advertises WPA2-PSK and FT-PSK; no other SSID is involved. The key server, an external RADIUS server or the cloud controller's key service, identifies keys and derives the FT hierarchy. The cloud management system delivers PMK-R0 to the primary AP and executes deletes; target APs obtain PMK-R1 on demand; and the PMS drives key removal.

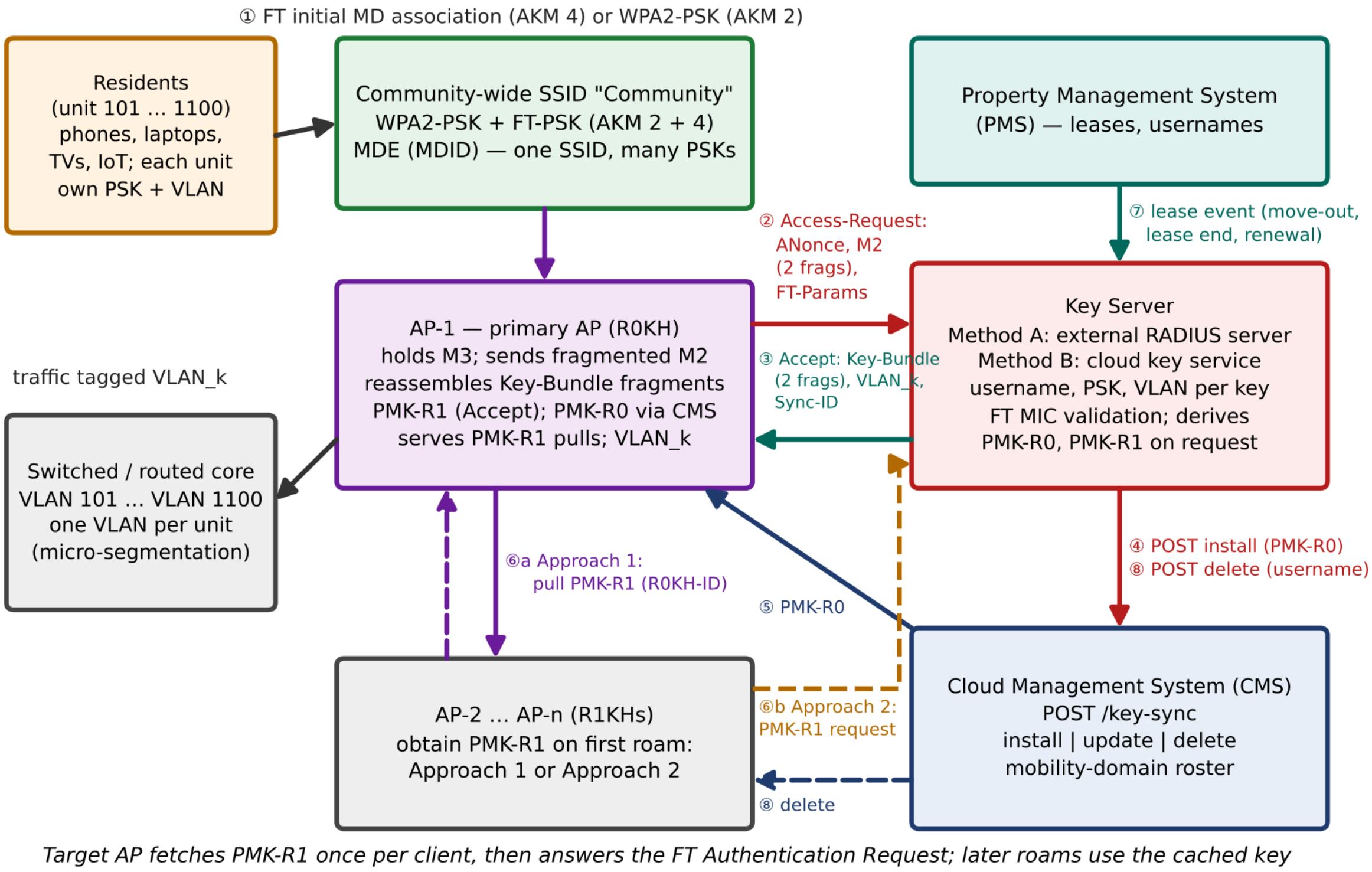

**Figure 2.** FT-mPSK architecture. AP-1 forwards the fragmented FT M2 (②); the key server identifies PSK_k and returns the Key-Bundle in two fragments with VLAN_k (③). It then posts PMK-R0 and the binding to the cloud management system (④), which pushes them to the primary AP (⑤). On a roam, the target AP obtains PMK-R1 from the primary AP (⑥a, Approach 1) or from the key server (⑥b, Approach 2). PMS lease events (⑦) trigger a delete by mPSK username through the same API (⑧).

### *4.1 Components*

- **Community-wide SSID.** WPA2-PSK with FT: AKM 2 (WPA2-PSK) and AKM 4 (FT-PSK) on 2.4 and 5 GHz, MDE with the property's MDID, FT over the air and over the DS. FT-capable clients use AKM 4, all others AKM 2, on the same SSID.
- **Primary AP (AP-1, R0KH).** The AP at which a client performs its FT initial mobility-domain association. It holds M3, sends the fragmented M2 to the key server, reassembles and verifies the Key-Bundle, completes the handshake with the PMK-R1 it contains, tags the client's traffic with VLAN_k, receives PMK-R0 from the cloud management system, and serves PMK-R1 pull requests from other APs (Approach 1).
- **R1KHs (AP-2 … AP-n).** All other APs of the mobility domain. When a client roams to one of them, it obtains the PMK-R1 derived for its own R1KH-ID, together with the client's binding ⟨username, VLAN_k, not-after⟩, from the primary AP or the key server, and caches it.
- **Key server.** In Method A, an external RADIUS server; in Method B, the key service of the cloud controller. It holds the PSK database with each key's VLAN, identity and lease end, the per-SSID PMK table, and the binding register. It identifies keys from M2, derives PMK-R0 and the serving AP's PMK-R1, builds the Key-Bundle, calls the Key-Sync API, and answers PMK-R1 requests (Approach 2).
- **Cloud management system (CMS).** Manages the APs and exposes POST /key-sync. It pushes PMK-R0 to primary APs and executes updates and deletes on all APs over their management sessions, and it provides each AP with the mobility-domain roster (R0KH-ID to AP address and AP keys). It relays sealed envelopes and never sees plaintext keys.

- **Property Management System (PMS).** Sends lease events (move-in, renewal, transfer, move-out, lease end) to the key server.

### *4.2 WLAN Profile Configuration*

**Table 4.** Recommended community-wide SSID configuration

| Parameter | Setting |
|---|---|
| SSID | One community-wide SSID, broadcast, on 2.4 and 5 GHz |
| Security | WPA2-PSK with mPSK; AKM 2 (PSK) and AKM 4 (FT-PSK); CCMP-128; PMF capable |
| 802.11r | Enabled; MDE with one MDID per mobility domain; FT over the air and over the DS |
| Key holders | AP of the FT initial association = R0KH (primary AP); every AP = R1KH |
| Key source | External RADIUS server (Method A) or cloud key service (Method B) |
| VLAN | Dynamic, per PSK (Tunnel-Private-Group-ID or cloud binding); client isolation between VLANs |
| Group-addressed traffic | Multicast-to-unicast conversion and proxy ARP/ND per VLAN |
| PMK-R1 delivery | On demand at the first roam to an AP: pull from the primary AP (Approach 1) or request to the key server (Approach 2) |
| Key distribution / revocation | Key-Sync API on the CMS: PMK-R0 install to the primary AP; update and delete by mPSK username on all APs |

### *4.3 Deployment Methods*

Both methods use the same key identification, Key-Bundle content, Key-Sync API, PMK-R1 approaches and lifecycle handling. They differ only in how the serving AP reaches the key server (Table 5).

**Table 5.** External RADIUS method versus cloud controller method

| Aspect | Method A: external RADIUS server | Method B: cloud controller |
|---|---|---|
| Key server | RADIUS server (on premises or hosted) | Key service of the cloud management system |
| AP → key server | Access-Request with fragmented M2 VSA | Validate request over the AP management session (TLS) |
| Key server → AP | Access-Accept with two-fragment Key-Bundle VSA and VLAN attributes | Validate response carrying the same Key-Bundle TLVs; no RADIUS size limit |
| Fragmentation | Required (Section 4.5) | Not required; the same TLV format is reused |
| PMK-R0 delivery | RADIUS server → POST /key-sync on the CMS → primary AP | Key service → POST /key-sync (internal) → primary AP |
| PMK-R1 at roam (Approach 2) | Access-Request with MPSK-R1-Request; Access-Accept with Key-Bundle (1 fragment) | PMK-R1 request over the AP management session |
| PMS integration | PMS → RADIUS server | PMS → cloud key service |
| Revocation | RADIUS server → POST /key-sync delete (username) | Key service → POST /key-sync delete (username) |

A cloud-hosted RADIUS service behaves like Method A. In Method B the transport has no attribute limit, but the Key-Bundle keeps the same TLV structure, so APs implement one parser for both methods.

### 4.4 RADIUS Attribute Dictionary (Method A)

Tables 6 and 7 list the attributes. The MPSK-EAPOL-M2 and MPSK-Key-Bundle VSAs use the fragment format of Section 4.5. Vendor-Type numbers are illustrative.

**Table 6.** Access-Request attributes (after M2, or for an Approach 2 PMK-R1 request)

| Attribute | Type | Length (octets) | Content / purpose |
|---|---|---|---|
| User-Name (1) / Calling-Station-Id (31) | Std | 12–17 / 17 | Client MAC (also a lookup hint) |
| Called-Station-Id (30) | Std | ≤ 50 | "BSSID:SSID" per RFC 3580 [6] |
| NAS-Identifier (32) / NAS-Port-Type (61) | Std | var. / 4 | AP identity / 19 = Wireless-IEEE-802.11 |
| Message-Authenticator (80) | Std | 16 | Mandatory [17], [18] |
| MPSK-ANonce (VSA 1) | VSA | 32 | ANonce from M1 |
| **MPSK-EAPOL-M2 (VSA 2), 1–2 fragments** | VSA | ≤ 255 each | Raw M2 (SNonce, MIC, RSNE, MDE, FTE); 121–125 octets for AKM 2, 239–290 for AKM 4 |
| MPSK-SSID / Client-MAC / BSSID / AP-MAC (VSA 3–6) | VSA | 1–32 / 6 / 6 / 6 | Inputs of Eqs. (1)–(9); AP identification for Key-Sync |
| **MPSK-FT-Params (VSA 7)** | VSA | 9–56 | MDID (2), R0KH-ID (1–48) and R1KH-ID (6) of the serving AP; present when the client selected an FT AKM |
| **MPSK-R1-Request (VSA 8)** | VSA | 25–72 | Approach 2 only: R0KH-ID from the FT Authentication Request, PMKR0Name (16), R1KH-ID of the requesting AP |

**Table 7.** Access-Accept attributes

| Attribute | Type | Length (octets) | Content / purpose |
|---|---|---|---|
| **MPSK-Key-Bundle (VSA 10), fragment 1** | VSA | 255 (value 247) | Header (Bundle-ID, M = 1, Index 0, Total 2, Enc-Alg) + E[0..242] |
| **MPSK-Key-Bundle (VSA 10), fragment 2** | VSA | ≤ 255 | Header (same Bundle-ID, M = 0, Index 1, Total 2) + rest of E |
| Tunnel-Type / Tunnel-Medium-Type / Tunnel-Private-Group-ID | Std | 6 / 6 / var. | VLAN (13) / IEEE-802 (6) / VLAN_k [5], [6] |
| Session-Timeout (27) | Std | 6 | Binding lifetime, never beyond the lease end |
| Message-Authenticator (80) | Std | 16 | Integrity of the whole Accept, including both fragments |

The Key-Bundle E is the encryption of the following TLVs: PSK_k, PMK_k, PMK-R1 for the serving AP's R1KH-ID, PMKR1Name, PMKR0Name, R0KH-ID, R1KH-ID, the mPSK username (Identity_k), key lifetime and a Sync-ID that correlates this exchange with the Key-Sync install. For AKM 2 clients the FT elements are omitted and the bundle fits in a single fragment. An Approach 2 response carries PMK-R1, PMKR1Name, PMKR0Name, the key-holder IDs, the username and the lifetime, and also fits in one fragment (Section 6.2). PMK-R0 is deliberately **not** in the Access-Accept: the serving AP does not need it to complete the handshake, and it is delivered by the cloud management system.

### 4.5 Fragmented VSA Format

Figure 3 shows the format. The Value of each fragmented VSA begins with a 4-octet header: **Bundle-ID** (1 octet, unique per message), **M | Index** (the More flag in the most significant bit and a 7-bit zero-based index), **Total** (number of fragments) and **Enc-Alg** (0 = none, used for M2; 1 = RFC 2548-style MD5 hiding; 2 = AES-256-GCM). The fragment payload is therefore at most 247 − 4 = 243 octets, and the largest attribute is 2 + 4 + 2 + 4 + 243 = 255 octets. A payload that fits in one fragment is sent with Index 0, Total 1 and M = 0, so one code path handles every case. The header maps one-to-one onto RFC 6929 Long Extended attributes [11], which gives a migration path to a standards-based encoding.

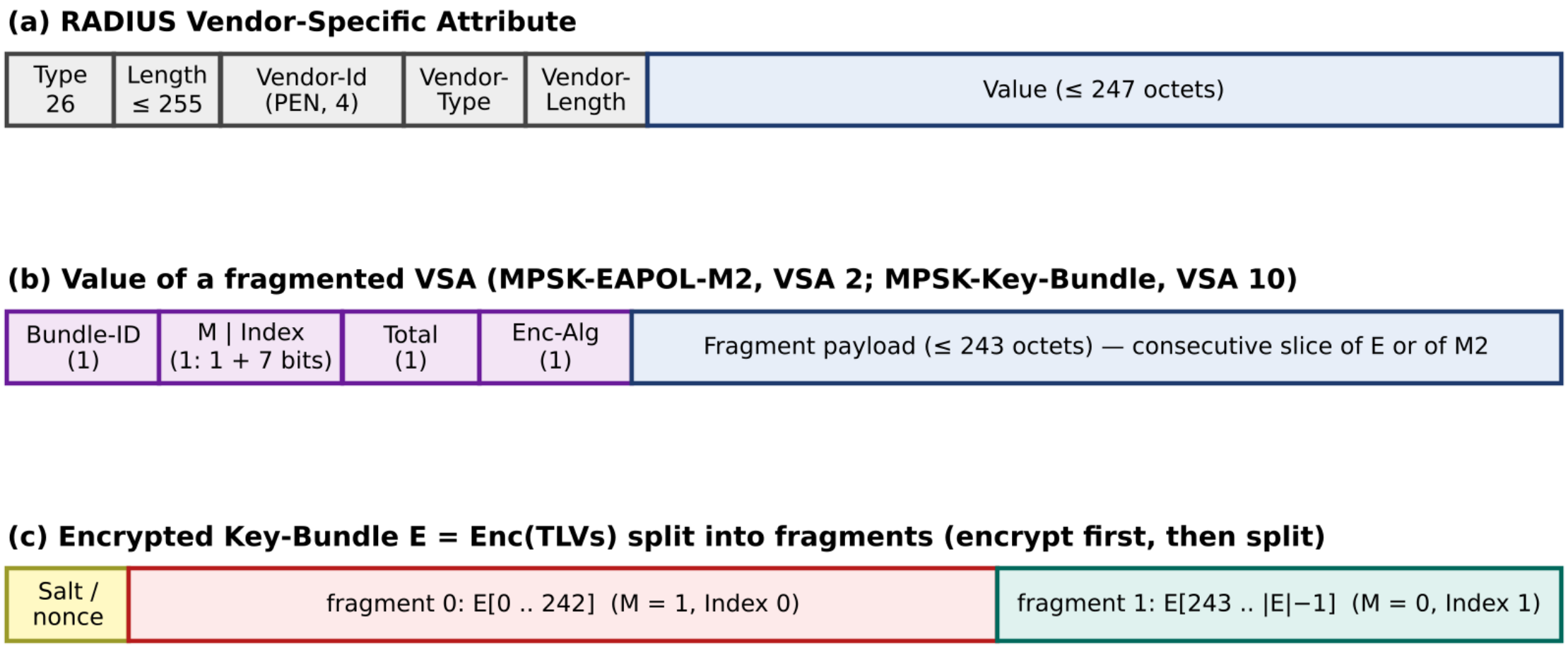

**Figure 3.** Fragmented VSA format: (a) RADIUS VSA; (b) 4-octet fragment header and payload, used for MPSK-EAPOL-M2 and MPSK-Key-Bundle; (c) the encrypted Key-Bundle is split only after encryption.

With RADIUS over UDP the Key-Bundle is protected with RFC 2548-style key hiding [13], extended with a 2-octet length prefix so that it can cover plaintexts longer than 255 octets. For plaintext $P = p_1 \parallel p_2 \parallel \ldots$ in 16-octet blocks:

$$b_1 = MD5(S \parallel RA \parallel Salt),\ c_1 = p_1 \oplus b_1;\ \ b_i = MD5(S \parallel c_(i-1)),\ c_i = p_i \oplus b_i \tag{11}$$

where S is the shared secret, RA the Request Authenticator and Salt a 2-octet value with its most significant bit set. The encrypted length for a plaintext of L octets is:

$$|E| = 2 + 16 \cdot \lceil (L + 2) / 16 \rceil,\quad n_frag = \lceil |E| / 243 \rceil \tag{12}$$

With RadSec or DTLS [14], [15] the bundle is sealed with AES-256-GCM [19] under a per-AP key (12-octet nonce, 16-octet tag). Encryption is always applied **before** fragmentation, so confidentiality and integrity cover the complete bundle. Both fragments travel in the same Access-Accept, so reassembly is stateless across RADIUS exchanges.

### *4.6 Key-Sync API on the Cloud Management System*

The cloud management system exposes one endpoint, **POST /api/v1/mpsk/key-sync**, used to install PMK-R0 at the primary AP and to update or delete AP-resident mPSK key material. The caller (the RADIUS server in Method A, the cloud key service in Method B) authenticates with mutual TLS 1.3 [20] and an OAuth 2.0 client-credentials token [21] bound to its certificate [22], with separate scopes for install/update and for delete. The body is JSON [23]. Tables 8 and 9 list the operations and fields.

**Table 8.** Key-Sync API operations

| Operation | Trigger | Targets | Effect on the AP |
|---|---|---|---|
| install | Successful key identification (FT initial association) | The primary AP (R0KH) only | Store PMK-R0, PMKR0Name and the binding ⟨username, PSK, PMK, VLAN, not-after⟩ so that it can serve PMK-R1 pulls |

| Operation | Trigger | Targets | Effect on the AP |
|---|---|---|---|
| update | Lease renewal or VLAN change | All APs ("all") | Change not-after or VLAN of every binding, PMK-R0 and PMK-R1 of the username |
| delete | PMS move-out, lease end, transfer, administrative revocation | All APs ("all") | Purge every PSK, PMK, PMK-R0, PMK-R1, PMKSA / PTKSA and binding of the username; deauthenticate every client bound to it |

**Table 9.** Key-Sync request fields

| Field | Content |
|---|---|
| request_id | UUID [24]; idempotency key. For install it equals the Sync-ID in the Key-Bundle. |
| op, issued_at, seq | install \| update \| delete; issue time; sequence number, strictly increasing per username |
| site_id, mdid, r0kh_id | Property, mobility domain, R0KH-ID of the primary AP (install) |
| subject | username (the mPSK / ePSK username) for every operation; plus client_mac for install. Deletes carry no MAC addresses. |
| targets | [{ap_id, env}] naming the primary AP for install, or "all" for update and delete |
| reason, not_after | Delete reason (lease_end, move_out, transfer, admin); new expiry for update |
| sig | Ed25519 signature [25] of the key server over the canonical body, verified by every AP |

The install envelope is sealed to the primary AP's device public key with HPKE [26] (DHKEM-X25519, AES-256-GCM), and the associated data binds it to this request, username, client and AP:

$$env_j = HPKE.Seal(pk_APj,\ info = \text{"FT-mPSK key-sync v1"},\ aad = request_id \parallel op \parallel username \parallel STA\text{-}ADDR \parallel ap_id_j \parallel seq,\ X_j) \tag{13}$$

where $X_j$ is the plaintext of Table 10. An envelope copied to another AP, client or request fails to open. Updates and deletes carry no keys; their authenticity comes from the server signature. Listing 1 shows an install and a delete. The CMS answers **202 Accepted** once it has queued the request and reports per-AP status (pending, applied, failed, offline) at a status URL. It rejects a malformed body (400), an unauthorised caller (401/403) and a sequence number not newer than the last applied (409). For offline APs it keeps the latest operation per username as desired state and delivers it before the AP serves clients again.

**Listing 1.** Key-Sync install after key identification, and delete after a PMS lease-end event.

```
POST /api/v1/mpsk/key-sync HTTP/1.1
Host: cms.example.net
Authorization: Bearer <certificate-bound access token>
Content-Type: application/json

{ "request_id": "7c1e0a52-4b8e-4f0e-9d0a-2f6c2b1d9e41",
  "op": "install", "issued_at": "2026-10-14T07:30:12Z", "seq": 3,
  "site_id": "community-0142", "mdid": "a1b2", "r0kh_id": "ap1.community-0142",
  "subject": { "username": "unit-1204", "client_mac": "02:5e:7a:11:42:9c" },
  "targets": [
    { "ap_id": "AP-1", "env": "<base64 HPKE: PMK-R0, PMKR0Name, binding>" } ],
  "sig": "<base64 Ed25519>" }

HTTP/1.1 202 Accepted
{ "request_id": "7c1e0a52-...", "status_url": "/api/v1/mpsk/key-sync/7c1e0a52-..." }

POST /api/v1/mpsk/key-sync HTTP/1.1
{ "request_id": "b9d4f1e0-2c37-4a51-8e0b-61f0d2a7c3aa",
  "op": "delete", "issued_at": "2026-11-30T23:59:00Z", "seq": 4,
  "site_id": "community-0142", "reason": "lease_end",
  "subject": { "username": "unit-1204" },
  "targets": "all", "sig": "<base64 Ed25519>" }
```

**Table 10.** Key-Sync install envelope plaintext X (primary AP; TLVs with a 2-octet header; worst-case octets)

| Element | AKM 4 | AKM 19 |
|---|---|---|
| PMK-R0, PMKR0Name | 34 + 18 | 50 + 18 |
| R0KH-ID (≤ 48) | 50 | 50 |
| PSK passphrase (≤ 63), PMK | 65 + 34 | 65 + 50 |
| mPSK username (≤ 64), VLAN, not-after | 66 + 4 + 10 | 66 + 4 + 10 |
| **Plaintext** | **281** | **313** |

## 5. Algorithm

Figure 4 shows the message exchange of an FT initial mobility-domain association through PMK-R0 delivery and the first roam.

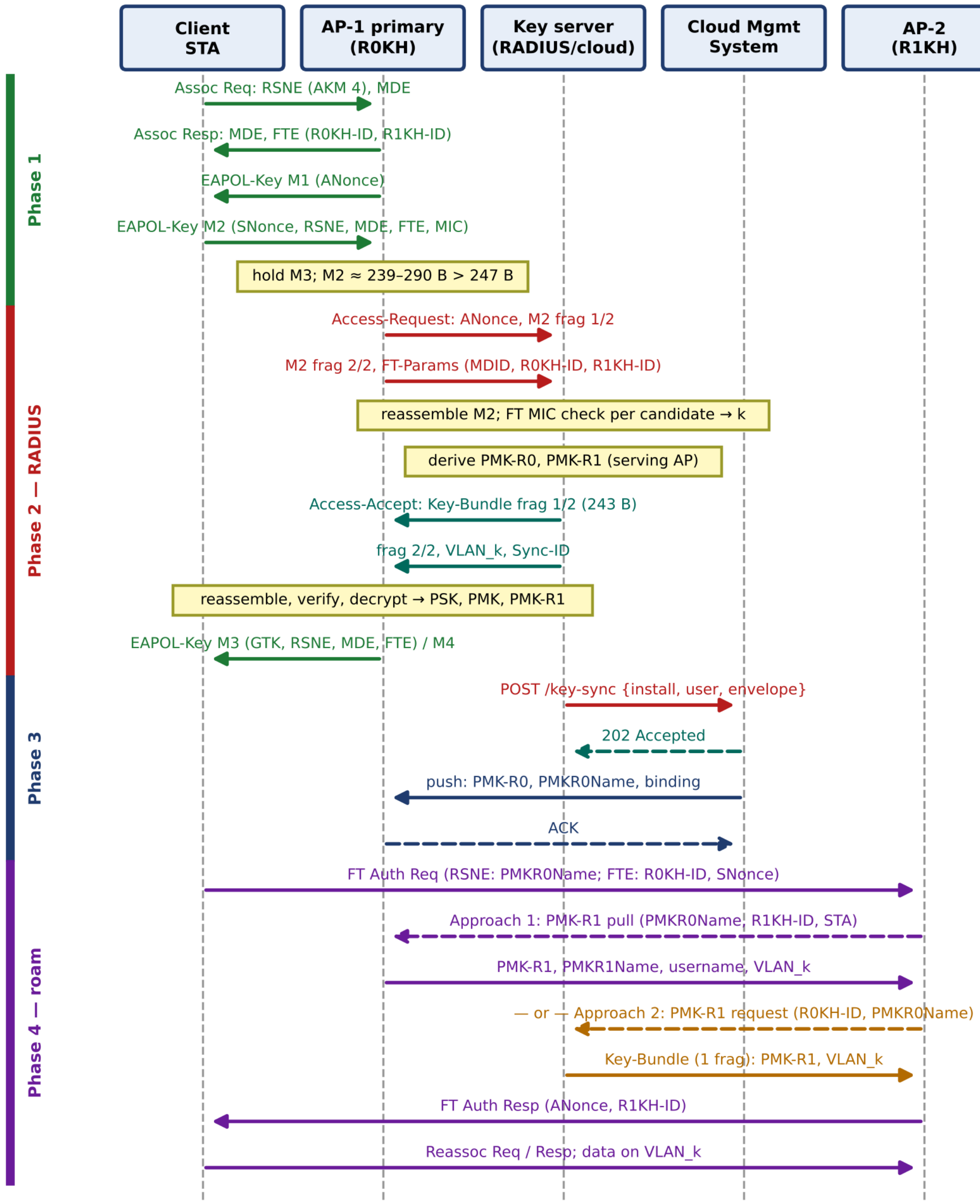

**Figure 4.** FT-mPSK message exchange (Method A): FT initial mobility-domain association on the community SSID, fragmented M2 in the Access-Request, key identification and FT derivation at the key server, two-fragment Key-Bundle in the Access-Accept, PMK-R0 delivered to the primary AP through the cloud management system, and a first roam in which the target AP obtains PMK-R1 by Approach 1 or Approach 2.

### *5.1 Phase 1: Association and Handshake on the Community SSID*

The client associates with AKM 4 and the MDE. AP-1 answers with the MDE and an FTE carrying its R0KH-ID and R1KH-ID, and sends M1 with ANonce. The client, which knows only its own unit's passphrase, derives PMK, PMK-R0, PMK-R1 and the PTK (Eqs. 1, 4–8), and returns M2 with SNonce, the RSNE, the MDE, the FTE and the MIC (Eq. 9). AP-1 does not know which key the client used, so it holds M3. A client that selects AKM 2 follows the same path with a plain M2.

### 5.2 Phase 2: Key Identification at the Key Server

1) AP-1 splits M2 into fragments of at most 243 octets and sends the Access-Request of Table 6 with ANonce, the M2 fragments, the client MAC, SSID, BSSID and MPSK-FT-Params.
2) The key server verifies Message-Authenticator, reassembles M2, and checks that the MDE and FTE in M2 match MPSK-FT-Params.
3) For AKM 4 it searches the candidate keys of the SSID (the key bound to the client MAC first, if any). For each candidate PMK_i it computes PMK-R0 and PMK-R1 for the serving AP (Eqs. 4, 6), derives the PTK (Eq. 8) and verifies the AES-CMAC MIC (Eq. 9); the first candidate whose MIC matches is PSK_k. It then computes PMKR0Name and PMKR1Name (Eqs. 5, 7) for the matched key only. For AKM 2 it evaluates Eqs. (2)–(3). Keys disabled by a PMS event are not candidates.
4) If no key matches, it returns Access-Reject and AP-1 deauthenticates the client and rate-limits the MAC.

### 5.3 Phase 2 (continued): Fragmented Key-Bundle and Completion of the Handshake

For the matched key k, the key server builds the Key-Bundle (Section 4.4), encrypts it as a whole (Eq. 11 or AES-GCM), splits the ciphertext into fragments of at most 243 octets (Eq. 12), and returns them in one Access-Accept with VLAN_k. AP-1 collects all MPSK-Key-Bundle VSAs and checks that they share one Bundle-ID, that their number equals Total (at most 2), that the indices run 0 … T−1 with the More flag set on all but the last, and that the reassembled length is valid for the encryption algorithm. It verifies the Response Authenticator and Message-Authenticator over the whole packet, and only then concatenates and decrypts. It sends M3 with the PTK derived from PMK-R1, receives M4, installs the keys and tags the client's traffic with VLAN_k. Any failure discards the bundle and deauthenticates the client (reason code 15); the client simply retries.

### 5.4 Phase 3: PMK-R0 to the Primary AP

After sending the Access-Accept, outside the handshake's critical path, the key server posts one install with request_id = Sync-ID. Its single envelope, sealed to the primary AP (Eq. 13), carries PMK-R0, PMKR0Name, the R0KH-ID and the binding ⟨username, PSK_k, PMK_k, VLAN_k, not-after⟩. The CMS pushes it to the primary AP, which verifies the signature and sequence number, installs PMK-R0 and the binding, and acknowledges; from then on it can serve PMK-R1 pull requests for the client. The key server records the binding (client MAC, username, R0KH-ID, PMKR0Name, VLAN, not-after) and erases PMK-R0. Because Eqs. (4)–(7) are deterministic, it can re-derive PMK-R0 or any PMK-R1 from PMK_k later.

### 5.5 Phase 4: Fast Roaming with Two Approaches for PMK-R1

When the client roams to AP-2, it sends an FT Authentication Request, over the air or over the DS, whose RSNE carries PMKR0Name and whose FTE carries its SNonce and the R0KH-ID of the primary AP. AP-2 computes the PMKR1Name it needs (Eq. 7). If it already holds that PMK-R1 from an earlier roam of the same client, it answers immediately. Otherwise it obtains PMK-R1 by one of two approaches before it sends the FT Authentication Response (Figure 5, Table 11).

- **Approach 1: pull from the primary AP (R0KH).** AP-2 reads the R0KH-ID, usually the MAC address of the primary AP, from the FTE of the FT Authentication Request and resolves it to the primary AP's address with the mobility-domain roster provided by the CMS. It sends a PMK-R1 pull request carrying a nonce, PMKR0Name, its own R1KH-ID and the client MAC. The primary AP finds PMK-R0 by PMKR0Name, derives PMK-R1 and PMKR1Name for AP-2 (Eqs. 6–7), and returns them with the client's username, VLAN_k and not-after. The exchange follows the 802.11r R0KH–R1KH pull model and is encrypted and authenticated with keys provisioned by the CMS.

- **Approach 2: request from the key server.** AP-2 sends the R0KH-ID, PMKR0Name, its R1KH-ID and the client MAC to the RADIUS server (Access-Request with MPSK-R1-Request, Method A) or to the cloud key service (Method B). The key server finds the binding by client MAC and PMKR0Name, checks that the username is active and that AP-2 belongs to the mobility domain, re-derives PMK-R0 from PMK_k and PMK-R1 for AP-2, and returns them in a Key-Bundle (one fragment) with VLAN_k.

AP-2 then derives the PTK (Eq. 8), sends the FT Authentication Response, completes the reassociation and tags the client's traffic with VLAN_k, so the unit's network follows the client around the property. It caches PMK-R1 and the binding until not-after or a delete. Approach 1 keeps the exchange on the property LAN and is the default. Approach 2 is used when the primary AP is unreachable or has not yet received PMK-R0, or when the operator does not permit key transfer between APs.

**Table 11.** The two approaches for delivering PMK-R1 to the target AP

| Aspect | Approach 1: pull from primary AP | Approach 2: request from key server |
|---|---|---|
| Source of PMK-R1 | Primary AP (R0KH), derived from PMK-R0 | RADIUS server or cloud key service, re-derived from PMK_k |
| How the source is found | R0KH-ID in the FTE of the FT Authentication Request, resolved with the CMS roster | The AP's configured key server |
| Request | Nonce, PMKR0Name, R1KH-ID, client MAC | R0KH-ID, PMKR0Name, R1KH-ID, client MAC (MPSK-R1-Request) |
| Response | PMK-R1, PMKR1Name, username, VLAN_k, not-after | Key-Bundle (1 fragment): PMK-R1, PMKR1Name, names, username, lifetime; VLAN_k |
| Protection | AP-to-AP keys provisioned by the CMS | RadSec / Message-Authenticator, or management-session TLS |
| Added latency (first roam to an AP) | One LAN round trip | One round trip to the key server |
| Depends on | Primary AP reachable and holding PMK-R0 | Key server reachable |
| Typical use | Default | Fallback; primary AP offline; no AP-to-AP key transfer |

A device that later performs a **new** FT initial association at another AP, for example after leaving the property and returning, goes through Phases 1–3 again. The client MAC lets the server try the bound key first, so the search costs one candidate. The new AP becomes the primary AP and receives a new PMK-R0; the previous PMK-R0 for that client expires at not-after or is removed by the next delete.

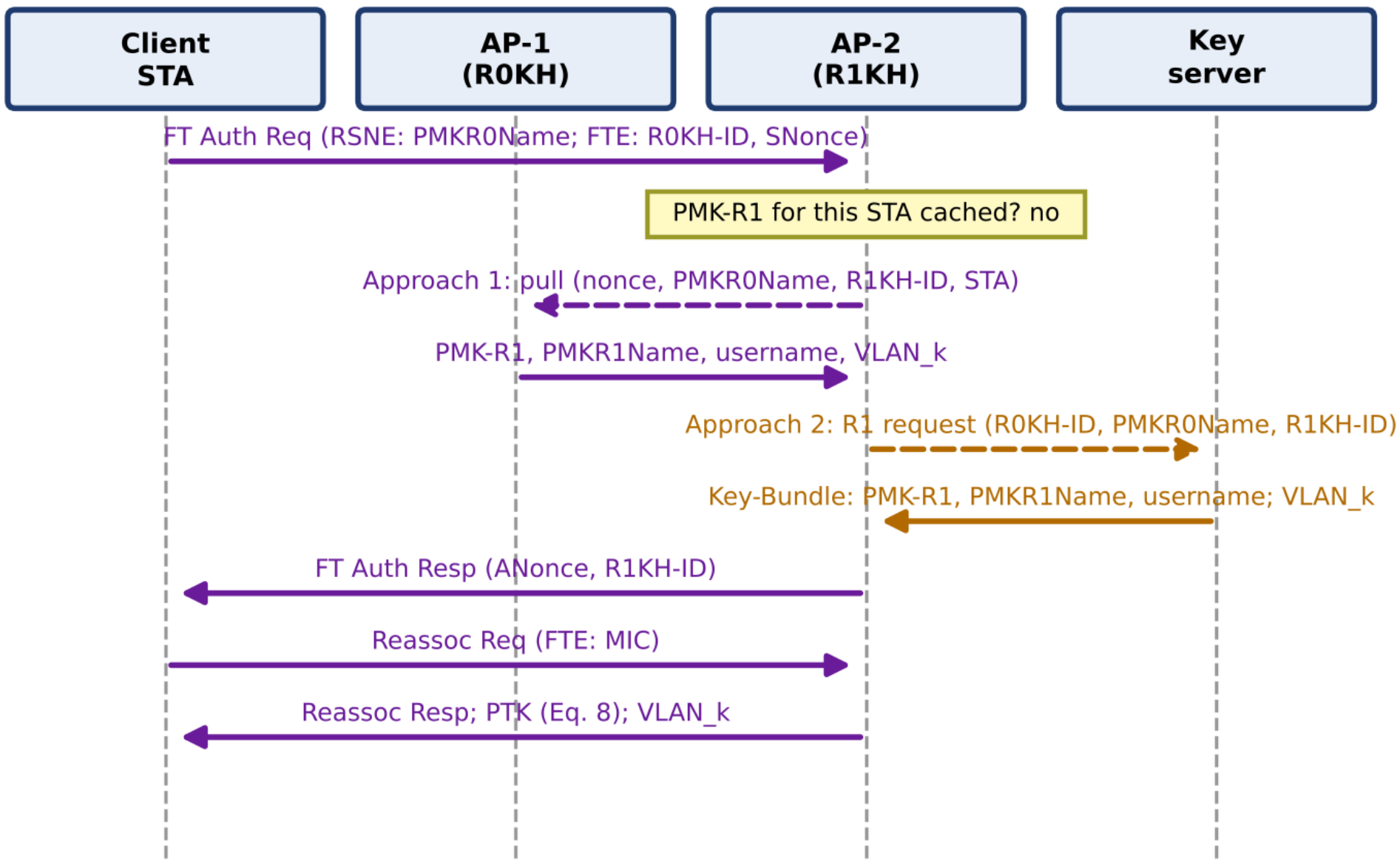

**Figure 5.** Fast roam from AP-1 to AP-2: AP-2 obtains PMK-R1 from the primary AP using the R0KH-ID in the FT Authentication Request (Approach 1) or from the key server (Approach 2), then answers the FT Authentication Request; the client keeps its unit's VLAN.

## *5.6 Method B: Cloud Controller*

In Method B, AP-1 sends a validate request with the same fields (ANonce, M2, client MAC, SSID, BSSID, FT parameters) to the cloud key service over its management session, and receives a validate response with the same Key-Bundle TLVs and VLAN_k. The session's TLS protects the bundle and there is no attribute limit, so no fragmentation is needed. The key service then calls the same POST /key-sync internally to deliver PMK-R0 to the primary AP, and it answers Approach 2 PMK-R1 requests over the same management sessions. Because M3 waits for this response, the cloud key service should run in a region close to the property; Section 7 quantifies the hold time.

## *5.7 Lease-Driven Key Lifecycle*

The PMS notifies the key server of lease events through an authenticated webhook, or the key server polls the PMS API. Before acting on a destructive event, the key server confirms it against the PMS lease record and applies it at the effective time stated by the PMS. Table 12 maps events to Key-Sync operations, and Figure 6 shows a move-out.

**Table 12.** PMS events and the resulting key server and Key-Sync actions

| PMS event | Key server action | Key-Sync operation | Result on APs and clients |
|---|---|---|---|
| Move-in / new lease | Create ⟨username, PSK, VLAN, lease end⟩; deliver PSK to resident | None until first association | PSK becomes an identification candidate |
| Renewal / extension | Update lease end | update {username, not_after} | Bindings and FT keys stay valid; no disconnection |
| Unit transfer | Retire old username and PSK; create username, PSK and VLAN for the new unit | delete {old username} | Old keys purged; resident joins with the new PSK |
| Move-out / lease end / eviction | Disable the username at the effective time | delete {username, targets: all} | Every PSK, PMK, PMK-R0 and PMK-R1 of the username purged on every AP; its devices deauthenticated (reason 2) |
| Administrative suspension | Disable the username | delete {username, targets: all} | As for move-out |

For a delete the key server proceeds in a fixed order. It first marks the username inactive, so that a reconnection attempt cannot succeed, because identification no longer finds a matching key. It then posts one delete for the **mPSK username** with targets = all; the request carries no client MAC addresses. Each AP finds every binding with that username and purges the PSK, PMK, PMK-R0, every PMK-R1 and any cached PMKSA or PTKSA associated with it, deauthenticates every associated client bound to that username with reason code 2 (previous authentication no longer valid), keeps the sequence number as a tombstone for the username, and acknowledges. Because the delete is keyed by username, it also removes the PMK-R1 values that APs obtained through Approach 1 or 2. An offline AP receives the delete from the CMS desired state before it serves clients again. Independently, every key installed on an AP carries a not-after time no later than the lease end, so an AP that never receives the delete still stops using the keys when the lease ends.

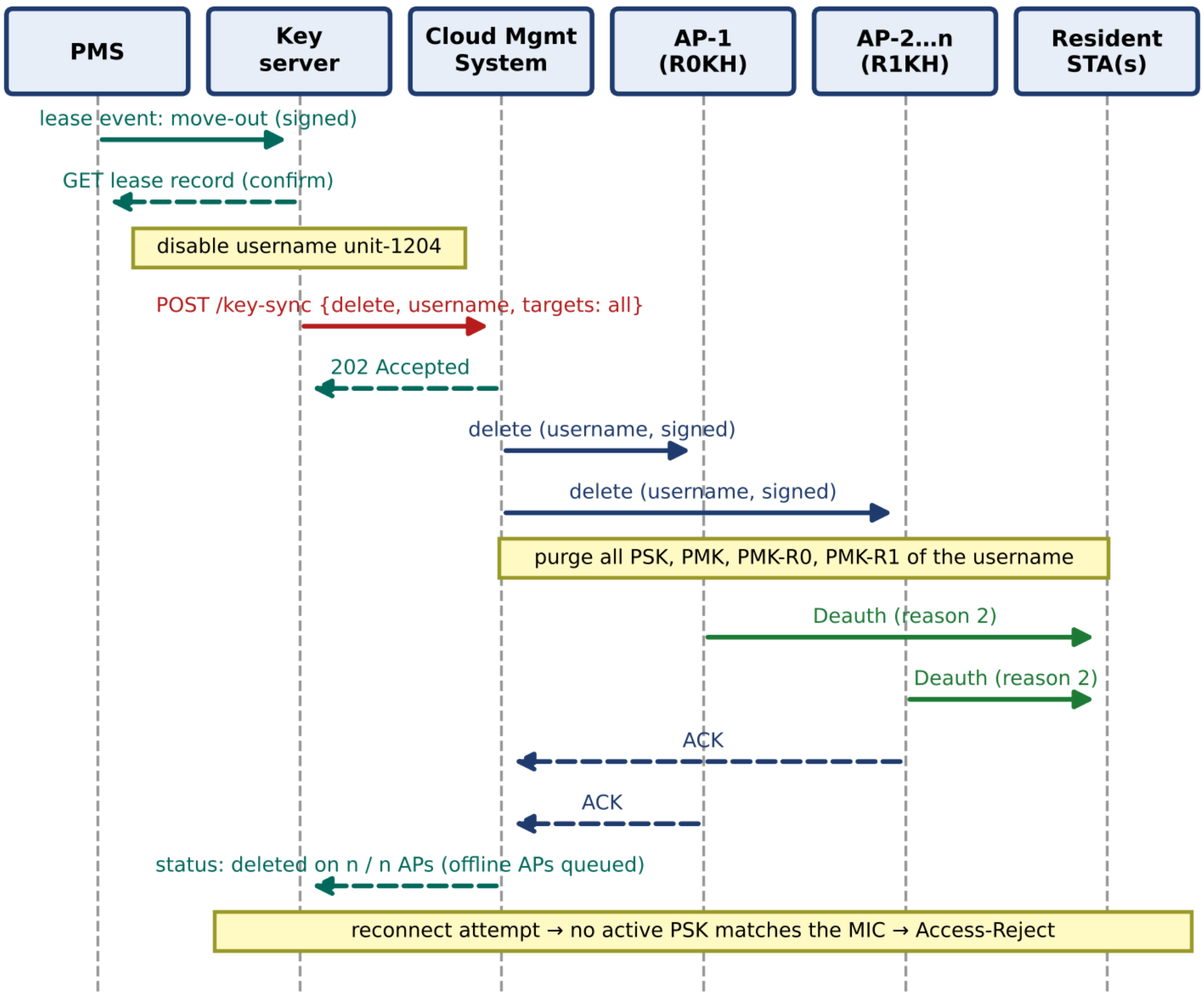

**Figure 6.** Lease-driven revocation: a PMS move-out event causes the key server to disable the username and post a Key-Sync delete for it; the CMS delivers it to every AP, which purges every PSK, PMK, PMK-R0 and PMK-R1 of that username and disconnects the resident's devices.

### *5.8 Algorithm Specification*

Figures 7 and 8 show the AP-side and key-server-side logic, and Algorithms 1–4 specify them.

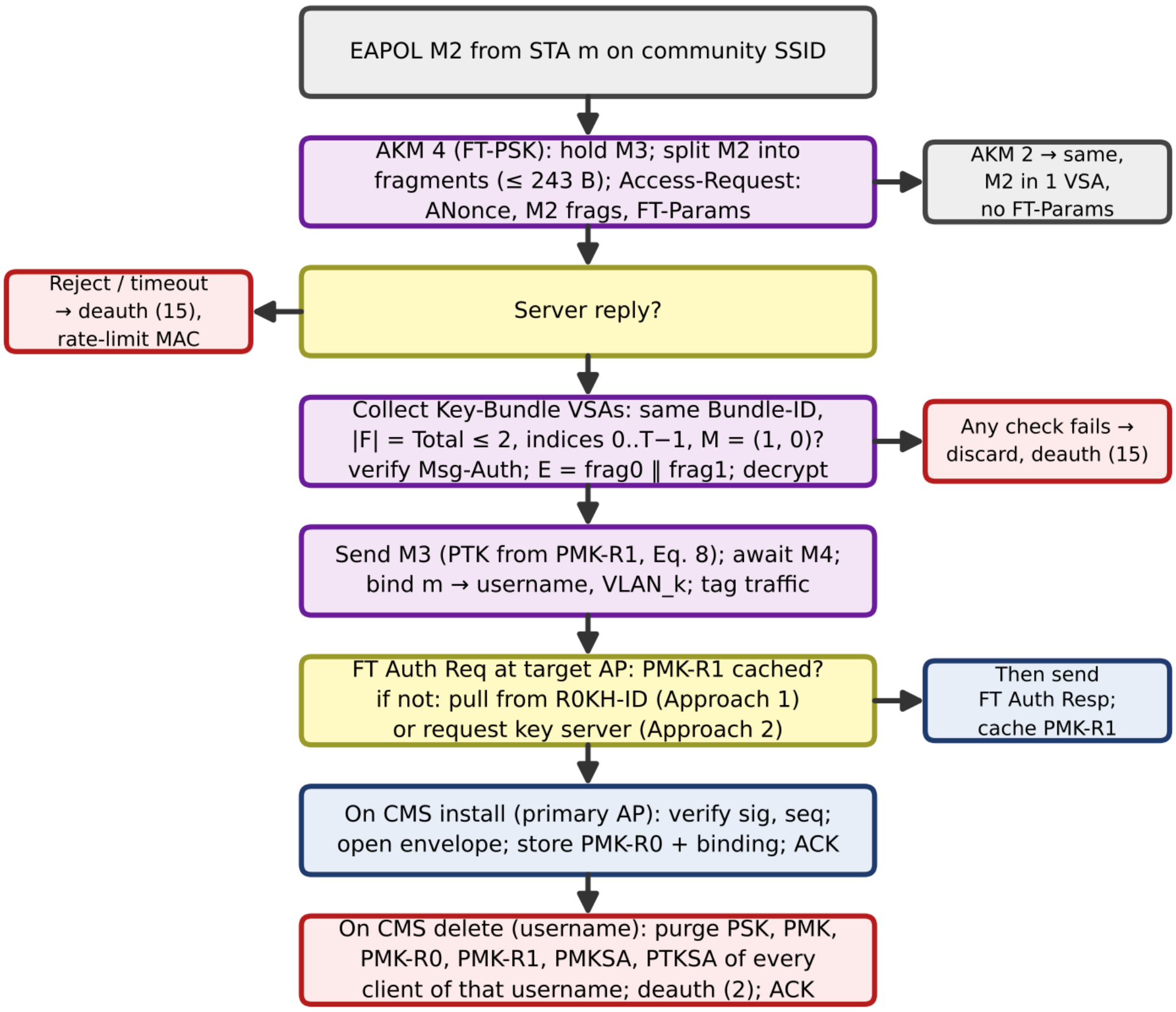

**Figure 7.** AP-side flow: fragmented M2, Key-Bundle reassembly and verification, completion of the FT handshake with PMK-R1, VLAN tagging, PMK-R1 fetch at roam time (Approach 1 or 2), and Key-Sync install and delete by username.

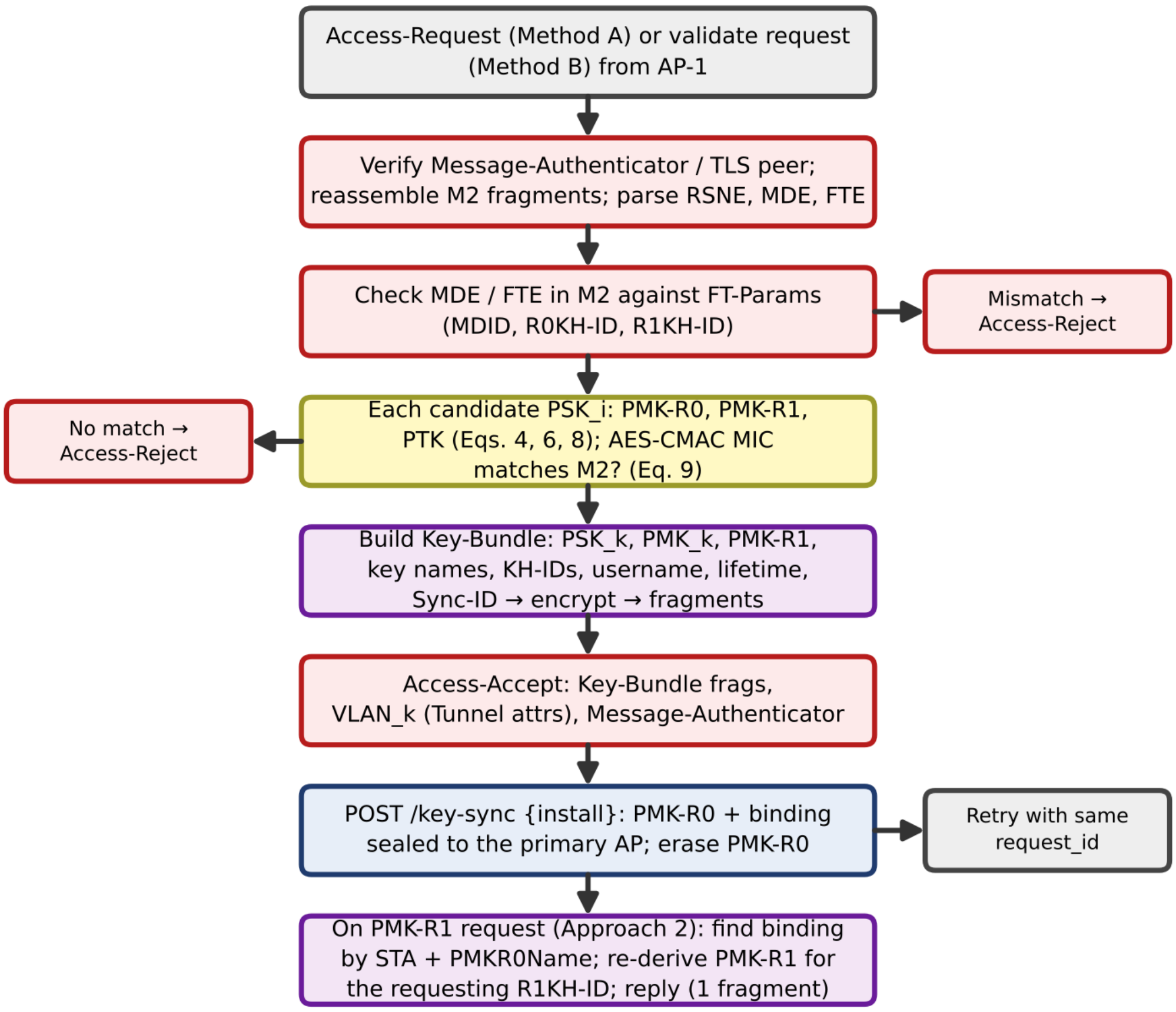

**Figure 8.** Key server flow: M2 reassembly, FT parameter check, FT MIC verification per candidate, Key-Bundle encryption and fragmentation, PMK-R0 install to the primary AP, and Approach 2 PMK-R1 requests.

**Algorithm 1.** FT-mPSK, AP side

```
Input : community SSID (AKM 2 + 4, MDID, R0KH-ID = AP.r0kh, R1KH-ID = AP.r1kh);
        device key pair (sk_AP, pk_AP); key server verification key pk_KS;
        mobility-domain roster from the CMS: R0KH-ID -> (AP address, AP key)
State : B[MAC] = (user, psk, pmk, vlan, not_after)
        R0[PMKR0Name] = (PMK-R0, MAC);  R1[PMKR1Name] = (PMK-R1, MAC)
        SEQ[user] = last applied seq (kept as tombstone after delete)

on EAPOL_M2(m, akm, frame):                         // FT initial MD association
    hold M3
    frags <- Fragment(frame, Enc-Alg = 0)               // Algorithm 3
    req   <- AccessRequest(ANonce, frags, SSID, m, BSSID, AP-MAC,
                           FT-Params = (MDID, AP.r0kh, AP.r1kh) if akm = 4,
                           Message-Authenticator)
    resp  <- send(KeyServer, req, timeout = T_rad, retries = 1)
    if resp is Reject or timeout: deauth(m, reason = 15); return
    E <- Reassemble(resp.KeyBundle[*]);  if E = ERROR: deauth(m, 15); return
    K <- Decrypt(E, secret, req.Authenticator)
    if akm = 4: PTK <- KDF(K.pmk_r1, "FT-PTK", SNonce | ANonce | BSSID | m)
                R1[K.r1name] <- (K.pmk_r1, m)
    else:       PTK <- PRF384(K.pmk, ...)
    send M3 (GTK, RSNE, MDE, FTE if FT) with PTK; await M4
    B[m] <- (K.user, K.psk, K.pmk, resp.VLAN, now + resp.TTL)
    tag traffic of m with resp.VLAN

on FT_Auth_Req(m, PMKR0Name, FTE):                  // roam (Figure 5)
    r1n <- Trunc128(SHA256("FT-R1N" | PMKR0Name | AP.r1kh | m))
    if R1[r1n] missing:
        X <- Pull(roster[FTE.R0KH-ID], nonce, PMKR0Name, AP.r1kh, m)  // Approach 1
        if X = ERROR or no reply within T_pull:
            X <- R1Request(KeyServer, FTE.R0KH-ID, PMKR0Name,
                           AP.r1kh, m)                                 // Approach 2
        if X = ERROR: send FT_Auth_Resp(m, status = failure); return
        R1[X.r1name] <- (X.pmk_r1, m)
```

```
        B[m] <- (X.user, -, -, X.vlan, X.not_after)
    send FT_Auth_Resp(m, ANonce, R1KH-ID); complete reassociation
    tag traffic of m with B[m].vlan

on Pull(nonce, PMKR0Name, r1kh, m) from peer AP:     // this AP is the R0KH
    require peer in roster and peer.r1kh = r1kh and R0[PMKR0Name] = (R0, m)
    R1  <- KDF(R0, "FT-R1", r1kh | m)
    R1N <- Trunc128(SHA256("FT-R1N" | PMKR0Name | r1kh | m))
    reply {nonce, R1, R1N, B[m].user, B[m].vlan, B[m].not_after}
          encrypted and authenticated with the AP-to-AP key

on KeySync(msg) pushed by CMS:                      // install | update | delete
    require Verify(pk_KS, msg.sig) and msg.seq > SEQ[msg.user]
    if msg.op = install:                            // this AP is the primary AP
        X <- HPKE.Open(sk_AP, msg.env, aad = msg.request_id | install
                       | msg.user | msg.mac | AP.id | msg.seq)
        R0[X.r0name] <- (X.pmk_r0, msg.mac)
        B[msg.mac]   <- (msg.user, X.psk, X.pmk, X.vlan, X.not_after)
    if msg.op = update:
        for each mac with B[mac].user = msg.user: apply msg.not_after / msg.vlan
    if msg.op = delete:                             // keyed by username only
        for each mac with B[mac].user = msg.user:
            purge B[mac], R0[* of mac], R1[* of mac], PMKSA[mac], PTKSA[mac]
            if associated(mac): deauth(mac, reason = 2)
    SEQ[msg.user] <- msg.seq;  ack(CMS, msg.request_id, AP.id, result)

every T_gc: purge all entries with not_after < now
```

**Algorithm 2.** FT-mPSK, key server side (Method A; Method B identical without fragments)

```
Input : DB D = {(user_i, PSK_i, VLAN_i, lease_end_i, active_i)};
        PMK[S][i] = PBKDF2(PSK_i, S); roster(MDID) -> {(ap_id, r1kh, pk_AP)}

on AccessRequest(req) with MPSK-EAPOL-M2:           // key identification
    verify Message-Authenticator (or TLS peer)
    M2 <- Reassemble(req.EAPOL-M2[*]);  (A, S, m, aa, ft) <- parse VSAs
    (akm, SN, MIC, mde, fte) <- parse(M2)
    if akm = 4: require mde.MDID = ft.MDID and fte.R0KH = ft.R0KH
                       and fte.R1KH = ft.R1KH
    for i in order(candidates(S), hint = binding(m)) with active_i:
        if akm = 4:
            (R0, salt) <- KDF(PMK[S][i], "FT-R0", S, ft.MDID, ft.R0KH, m)
            R1  <- KDF(R0, "FT-R1", ft.R1KH | m)
            PTK <- KDF(R1, "FT-PTK", SN | A | ft.BSSID | m)
            if AES_CMAC(PTK.KCK, zero_mic(M2)) == MIC: k <- i; break
        else:
            PTK <- PRF384(PMK[S][i], aa, m, A, SN)
            if HMAC_SHA1_128(PTK.KCK, zero_mic(M2)) == MIC: k <- i; break
    if no k: return AccessReject
    na <- min(lease_end_k, now + T_policy);  sid <- UUID()
    P  <- TLVs(PSK_k, PMK[S][k], user_k, lifetime = na - now, sid)
    if akm = 4:
        R0N <- Trunc128(SHA256("FT-R0N" | salt))
        R1N <- Trunc128(SHA256("FT-R1N" | R0N | ft.R1KH | m))
        P   <- P | TLVs(R1, R1N, R0N, ft.R0KH, ft.R1KH)
    E  <- Encrypt(P, secret, req.Authenticator)        // Eq. (11) or AES-GCM
    reply AccessAccept(Fragment(E), VLAN_k, Session-Timeout, Message-Authenticator)
    if akm = 4:                                        // after the reply
        seq <- next_seq(user_k);  ap1 <- roster entry with r1kh = ft.R1KH
        X   <- {R0, R0N, ft.R0KH, PSK_k, PMK[S][k], VLAN_k, na}
        env <- HPKE.Seal(ap1.pk, X, aad = sid | install | user_k | m | ap1.id | seq)
        POST /key-sync ({request_id = sid, op = install, user = user_k, mac = m,
                         seq, targets = [(ap1.id, env)]}, sig = Sign(sk_KS, ...))
        record binding(m -> user_k, VLAN_k, ft.R0KH, R0N, na);  erase R0

on AccessRequest(req) with MPSK-R1-Request:         // Approach 2
    verify Message-Authenticator;  (r0kh, R0N, r1kh) <- parse;  m <- Calling-Station-Id
    b <- binding(m) with b.R0N = R0N and b.r0kh = r0kh and active(b.user)
    require b and r1kh in roster(MDID) and r1kh matches the requesting NAS
    (R0, -) <- KDF(PMK[S][b.key], "FT-R0", S, MDID, r0kh, m)
    R1  <- KDF(R0, "FT-R1", r1kh | m)
    R1N <- Trunc128(SHA256("FT-R1N" | R0N | r1kh | m))
    E   <- Encrypt(TLVs(R1, R1N, R0N, r0kh, r1kh, b.user, lifetime))
    reply AccessAccept(Fragment(E), VLAN_b, Message-Authenticator);  erase R0, R1
```

**Algorithm 3.** Fragmentation and reassembly (both directions)

```
Fragment(E, alg):                                         // sender
    id <- random_octet();  T <- ceil(|E| / 243)
    for j in 0..T-1:
        chunk <- E[243*j .. min(243*(j+1), |E|) - 1]
        emit VSA(PEN, type, id | (M = (j < T-1)) << 7 | j | T | alg | chunk)

Reassemble(vsas):                                         // receiver
    F <- [v for v in vsas if v.vendor = PEN and v.type = type]
    require all v.id equal;  T <- F[0].total;  require |F| = T and T <= 2
    sort F by index;  require indices = 0..T-1 and M-flags = (1,..,1,0)
    E <- F[0].chunk || ... || F[T-1].chunk
    if alg = 1: require |E| >= 18 and (|E| - 2) mod 16 = 0    // salt + blocks
    return E    // decrypt only after Response Authenticator and
                // Message-Authenticator verify the full packet
```

**Algorithm 4.** PMS event handling at the key server

```
on PMSEvent(ev):          // move_in | renewal | transfer | move_out | lease_end
    require VerifyWebhook(ev) and ev consistent with PMS.GET(lease(ev.lease_id))
    wait until ev.effective_time
    u <- username_of(ev.unit, ev.resident)
    switch ev.type:
      move_in:   create (user = u, PSK, VLAN, lease_end, active = true)
      renewal:   lease_end_u <- ev.new_end
                 POST /key-sync {op = update, user = u, targets = all,
                                 not_after = min(ev.new_end, now + T_policy)}
      transfer, move_out, lease_end:
                 active_u <- false                      // blocks reconnection first
                 POST /key-sync {op = delete, user = u, targets = all,
                                 seq = next_seq(u), reason = ev.type}
                 if ev.type = transfer: create username, key and VLAN for new unit
    poll status until applied on all APs; alert on failed or offline APs
```

## 6. Size Budget

### 6.1 Access-Request: Raw M2

Table 13 gives the size of the raw M2. The EAPOL header and key descriptor take 99 octets (with a 16-octet MIC). For AKM 4 the key data adds the RSNE (including its PMKID list), the MDE and the FTE, whose R0KH-ID sub-element is 1–48 octets. The FT M2 therefore needs two fragments except in the rare case of a very short R0KH-ID without PMF.

**Table 13.** Raw EAPOL-Key M2 size (octets)

| Element | WPA2-PSK (AKM 2) | FT-PSK (AKM 4) |
|---|---|---|
| EAPOL header + key descriptor (MIC 16) | 99 | 99 |
| RSNE (without / with PMF group cipher) | 22 / 26 | 40 / 44 (with PMKID list) |
| MDE | — | 5 |
| FTE (MIC control, MIC, ANonce, SNonce, R1KH-ID, R0KH-ID 1–48) | — | 95–142 (100 with 6-octet R0KH-ID) |
| **Total** | **121–125** | **239–290 (244–248 typical)** |
| Fragments (payload ≤ 243) | 1 | 2 (1 only if ≤ 243) |

### 6.2 Access-Accept: Key-Bundle

Table 14 gives the Key-Bundle budget with Eq. (12). A bundle fits in one fragment only if its plaintext is at most 238 octets. With FT, the PSK (up to 63 characters), the R0KH-ID (up to 48 octets) and the identity (up to 64 octets) make that untrue for typical and worst-case configurations, so the AP and key server **must** support fragmentation. With short values a SHA-256 bundle can fit in one fragment; the format handles both cases with one code path (Total = 1 or 2).

**Table 14.** Key-Bundle size budget (TLVs with 2-octet header; octets)

| Element | AKM 2 | AKM 4 typical* | AKM 4 worst | AKM 19 worst |
|---|---|---|---|---|
| PSK passphrase | 65 | 65 | 65 | 65 |
| PMK | 34 | 34 | 34 | 50 |
| PMK-R1 (serving AP) | — | 34 | 34 | 50 |
| PMKR1Name + PMKR0Name | — | 36 | 36 | 36 |
| R0KH-ID + R1KH-ID | — | 8 + 8 | 50 + 8 | 50 + 8 |
| Identity | 66 | 34 | 66 | 66 |
| Lifetime + Sync-ID | 6 + 18 | 6 + 18 | 6 + 18 | 6 + 18 |
| **Plaintext L** | **189** | **243** | **317** | **349** |
| **Encrypted \|E\| (Eq. 12)** | **194** | **258** | **322** | **354** |
| **Fragments (payload split)** | **1** | **2** (243 + 15) | **2** (243 + 79) | **2** (243 + 111) |

*Typical: 63-character PSK, 6-octet R0KH-ID, 32-character identity. With AES-256-GCM over RadSec the encrypted size is L + 28 octets (271–377 for FT), with the same fragment counts. An Approach 2 response carries PMK-R1, PMKR1Name, PMKR0Name, R0KH-ID, R1KH-ID, the username and the lifetime: 200 octets of plaintext (216 for AKM 19), or 210 (226) octets encrypted, which fits in one fragment.

### *6.3 Key-Sync Payload*

The install carries one envelope, for the primary AP, with 281 octets of plaintext for AKM 4 (313 for AKM 19; Table 10). HPKE adds 48 octets (32-octet encapsulated key, 16-octet tag) and base64 adds a third, so the whole install request is about 1 KB, independent of the size of the mobility domain (Table 15). Update and delete requests carry only the username and are a few hundred octets.

**Table 15.** Key-Sync payload sizes (worst case, octets)

| Item | Plaintext<br>AKM 4 / AKM 19 | Sealed (HPKE)<br>AKM 4 / AKM 19 | In JSON |
|---|---|---|---|
| Install envelope (primary AP) | 281 / 313 | 329 / 361 | 440 / 484 (base64) |
| Install request | — | — | ≈ 1 KB |
| Update or delete request (any n) | — | — | ≈ 300 |

## 7. Computational Complexity and Latency

The expensive part of PSK handling, PBKDF2 with 4,096 iterations (Eq. 1), is pre-computed once per key and SSID, so identification works on a PMK table. Table 16 compares the per-candidate cost. For FT-PSK each candidate needs the R0 KDF (two HMAC-SHA256 operations for 384 bits of output), the R1 KDF (one), the PTK KDF (two) and one AES-CMAC, compared with four HMAC-SHA1 operations for plain WPA2-PSK. The R0 and R1 steps cannot be pre-computed, because they depend on the client MAC and the serving AP. With N keys per SSID the search takes at most N candidates and N/2 on average; the MAC hint reduces it to one candidate for known devices, and scoping candidates to the property keeps N small.

**Table 16.** Per-candidate identification cost

| Mode | Per candidate |
|---|---|
| WPA2-PSK (AKM 2) | PRF-384 (3 HMAC-SHA1) + MIC (1 HMAC-SHA1) |

| Mode | Per candidate |
|---|---|
| FT-PSK (AKM 4) | R0 KDF (2 HMAC-SHA256) + R1 KDF (1 HMAC-SHA256) + PTK KDF (2 HMAC-SHA256) + AES-CMAC |

The client waits for M3 while AP-1 holds it, so the hold time must stay well within the client's handshake timeout:

$$T_hold = RTT(AP \leftrightarrow key\ server) + T_search(N) + T_frag + T_decrypt \qquad (14)$$

Fragmentation and reassembly (T_frag) are constant-time copies. The Key-Sync install is sent after the Access-Accept, so it is not part of T_hold. The first roam of a client to a given AP adds the time to fetch PMK-R1:

$$T_fetch = RTT(AP_j \leftrightarrow R0KH) + T_KDF\ \ (Approach\ 1);\quad T_fetch = RTT(AP_j \leftrightarrow key\ server) + T_KDF\ \ (Approach\ 2) \qquad (15)$$

T_KDF is one or two HMAC-SHA256 operations. Approach 1 normally has the lower T_fetch, because the primary AP is on the property LAN; Approach 2 adds the round trip to the key server but does not depend on the primary AP. The target AP caches PMK-R1, so later roams of the same client to the same AP cost one FT authentication and one reassociation exchange (four frames) with no fetch. Revocation latency is:

$$T_revoke = T_PMS{\rightarrow}KS + T_POST + max_j\ (\ T_push,j + T_apply,j\ ) + T_deauth \qquad (16)$$

bounded, for an offline AP, by the earlier of its reconnection to the CMS and the not-after time, which is never later than the lease end.

**Table 17.** Per-client cost summary

| Operation | When | Cost |
|---|---|---|
| Key identification | FT initial association (first join, new primary AP, expiry) | ≤ N × (5 HMAC-SHA256 + AES-CMAC) on server; 1 round trip |
| M2 fragmentation / Key-Bundle reassembly | Every identification (Method A) | 2 + 2 VSAs; O(1) |
| Key-Bundle encryption | Every Accept | ≤ 22 MD5 blocks or 1 AES-GCM |
| Key-Sync install | Once per identification | 1 POST (≈ 1 KB) to the primary AP |
| PMK-R1 fetch, Approach 1 | First roam of a client to an AP | 1 AP-to-AP exchange; 1 KDF at the primary AP |
| PMK-R1 fetch, Approach 2 | First roam (primary AP unavailable) | 1 key-server round trip; R0 + R1 KDF at the server |
| FT roam with cached PMK-R1 | Later roams to the same AP | FT auth + reassociation (4 frames); no fetch |
| Key-Sync delete | Move-out, lease end, transfer | 1 POST (≈ 0.3 KB) per username; purge + deauth on every AP |

## 8. Security Analysis

### 8.1 What mPSK Adds to WPA2-PSK Security

In a single-PSK network, anyone who knows the passphrase and captures a client's four-way handshake can derive that client's PTK. With mPSK each unit's PMK is different, so a resident who knows their own key cannot decrypt another unit's unicast traffic. Per-PSK VLANs extend this isolation to layer 2 and 3. WPA2-PSK remains exposed to offline dictionary attacks on captured handshakes. Each key should therefore be randomly generated with high entropy (at least 20 characters), rotated on lease events, and never reused across units. PMF should be enabled where clients support it.

### 8.2 Group-Addressed Traffic across Units

All clients of a BSS share the GTK regardless of their PSK. A resident's device can therefore decrypt group-addressed frames sent on the BSS and, in principle, inject group-addressed frames. To keep units isolated, APs should not forward one VLAN's group-addressed traffic onto the air for other units: they should convert multicast to unicast per VLAN, answer ARP and neighbour discovery by proxy, and drop client-originated group frames that do not belong to the sender's VLAN. This is independent of FT, but it is part of making per-PSK micro-segmentation hold.

### 8.3 RADIUS Transport and Fragment Integrity

Message-Authenticator is mandatory on every Access-Request and Access-Accept in light of Blast-RADIUS [18], [17], and RadSec or DTLS with AES-GCM is strongly preferred over MD5 key hiding, which rests on the shared secret. The Key-Bundle is encrypted as one unit before fragmentation, so splitting reveals nothing beyond the total length. Both fragments travel inside one authenticated Access-Accept. Dropping, reordering, duplicating or injecting fragments breaks the Response Authenticator and Message-Authenticator, the GCM tag, or the structural checks of Algorithm 3, and every failure ends in a clean deauthentication, never in the use of a partial key. The M4 MIC, computed by the client from its own PMK-R1, confirms that the returned keys belong to this handshake.

### 8.4 Key Placement and the Cloud Management System

The key server is the only party that derives PSK-rooted FT keys at identification and for Approach 2, and it keeps the key names, not PMK-R0, afterwards. PMK-R0 is delivered only to the primary AP, the R0KH named in R0KH-ID, inside an HPKE envelope sealed to that AP's device key (Eq. 13), so a compromised CMS cannot read it, redirect it or reuse it for another client. A target AP holds a PMK-R1 only for clients that actually roamed to it, derived for its own R1KH-ID, which keeps fewer keys on fewer APs than a proactive push. Approach 1 pull messages between APs must be encrypted and authenticated, for example with pairwise AP keys provisioned by the CMS, and the primary AP serves a pull only to an AP listed in the mobility-domain roster whose identity matches the requested R1KH-ID. Approach 2 requests are authenticated RADIUS (Message-Authenticator, preferably RadSec) or management-session requests, and the key server serves PMK-R1 only for an active username, for an AP of the same mobility domain, and only when the requesting NAS matches the R1KH-ID. Every Key-Sync instruction is signed by the key server, so the CMS cannot forge installs or deletes; it can delay or drop them, and the not-after bound limits the effect of a dropped delete. Because Eqs. (4)–(7) are deterministic, keys derived by the key server on behalf of the R0KH are identical to those the R0KH would produce.

### 8.5 API Authorisation, Replay and PMS Integrity

The API requires mutual TLS 1.3 and certificate-bound OAuth 2.0 tokens [22]. Delete is a separate scope from install and update, so a credential that can only provision cannot revoke. Replay is prevented at three levels: request_id makes retries idempotent at the CMS; the per-username sequence number makes each AP reject any instruction not newer than the last applied; and the tombstone kept for the username after a delete makes a replayed install for a revoked key fail even if it is signed and correctly sealed. A forged move-out is a denial-of-service attack on a resident, so PMS webhooks are authenticated, the key server confirms each destructive event against the PMS record before acting, and bulk deletions above an operator threshold require confirmation.

### 8.6 Completeness of Revocation

Revocation acts on three layers. Disabling the username in the key server stops new identification and Approach 2 requests immediately. The Key-Sync delete, keyed by username, removes all AP-resident key material of that username, including PMK-R1 values fetched by Approach 1 or 2, and disconnects its devices within T_revoke (Eq. 16). The not-after bound guarantees expiry at the lease end even for an AP that cannot be reached. A former resident's device still has the passphrase configured, but no active key matches its MIC, so it cannot join again.

**Table 18.** Threat and mitigation summary

| Threat | Impact | Mitigation |
|---|---|---|
| Offline dictionary attack on a captured handshake | One unit's PSK exposed | High-entropy per-unit keys; rotation on lease events |
| Resident decrypting another unit's unicast traffic | Privacy loss | Different PMK per unit (mPSK); per-PSK VLAN |
| Cross-unit group-addressed traffic | Discovery or leakage across units | Multicast-to-unicast per VLAN; proxy ARP/ND; group-frame filtering |
| Eavesdropping on RADIUS | PSK / PMK / PMK-R1 exposure | RadSec / DTLS; AES-GCM; Message-Authenticator |
| Fragment drop, reorder or injection | Bundle rejected | Single-packet reassembly; integrity before decryption |
| CMS compromise | Exposure or misuse of keys | Per-AP HPKE envelopes; key-server signatures |
| Forged PMK-R1 pull or request | PMK-R1 disclosed to a rogue AP | Encrypted, authenticated AP-to-AP channel; roster check; RADIUS / RadSec authorisation of the requesting NAS |
| Replayed install after delete | Revoked key reinstalled | Per-username seq; tombstones; request_id idempotency |
| Forged PMS event or forged delete | Wrongful revocation | Authenticated webhook; PMS confirmation; signed instructions; separate delete scope |
| R1KH (AP) compromise | PMK-R1 of clients that roamed to it | PMK-R0 only at primary AP; on-demand per-AP PMK-R1; not-after |
| AP offline during delete | Stale keys on one AP | Desired-state delivery on reconnect; not-after ≤ lease end |

## 9. Experimental Evaluation

### 9.1 Testbed

The testbed comprised 4 commercial enterprise APs managed by a cloud management system forming one mobility domain on a community-wide SSID with AKM 2 and AKM 4, an external RADIUS server for Method A, the cloud key service for Method B, a Key-Sync endpoint on the cloud management system , and a PMS (simulated by sending messages to RADIUS) provisioned with 1,000 units, each with its own PSK and VLAN. FT over the air and over the DS was enabled. RADIUS and API traffic was captured to verify attribute lengths, fragment layout and request contents, and over-the-air traffic was captured with an additional AP in sniffer mode.

### 9.2 Methodology

1) **TC1 Fragmented M2:** an FT-PSK client associates; the Access-Request carries M2 in the number of fragments predicted by Table 13, and the server reassembles it.
2) **TC2 Fragmented Key-Bundle:** the Access-Accept carries the Key-Bundle in two fragments; the AP reassembles, verifies and decrypts it; the key names match the server log.

3) **TC3 FT initial association and VLAN:** the client completes the four-way handshake with the server-supplied PMK-R1 and is placed in its unit's VLAN.
4) **TC4 Key-Sync install:** PMK-R0 and the binding are present only on the primary AP; T_POST and T_push are recorded.
5) **TC5 FT roam, Approach 1:** the client roams AP-1 → AP-2 → AP-3, over the air and over the DS; each target AP pulls PMK-R1 from the primary AP using the R0KH-ID in the FT Authentication Request; the VLAN does not change; T_fetch is recorded.
6) **TC6 FT roam, Approach 2:** as TC5 with the target AP obtaining PMK-R1 from the key server, including with the primary AP offline; T_fetch is recorded.
7) **TC7 Mixed clients:** AKM 2 clients on the same SSID are identified and placed in the correct VLAN.
8) **TC8 Lease end:** a PMS move-out triggers a Key-Sync delete for the username; PSK, PMK, PMK-R0 and every PMK-R1 of the username are absent from every AP, including APs that fetched PMK-R1, all of its devices are deauthenticated, and reconnection is rejected; T_revoke is recorded.
9) **TC9 Lease renewal:** a PMS renewal triggers an update; not-after changes on every AP and no device is disconnected.
10) **TC10 Robustness:** a dropped or corrupted Key-Bundle fragment leads to rejection and a clean retry; an AP offline during a delete applies it on reconnection; a replayed install after a delete and a forged PMK-R1 pull request are rejected.
11) **TC11 Method B:** TC3–TC9 repeated with the cloud controller method.

### *9.3 Results*

In an initial implementation that returned the PSK and the FT key material in a single attribute, the payload exceeded the 255-octet RADIUS attribute limit and the Access-Accept could not be encoded, which is the size problem quantified in Tables 13 and 14. With two-fragment VSAs in both directions the AP reassembled and decrypted the Key-Bundle and completed the FT initial mobility-domain association. Table 19 records interoperability by client category, and Table 20 the system measurements. Measured values to be added.

**Table 19.** Interoperability results by client category

| Client category | Example device / OS | TC3 | TC5 | TC6 | TC7 | TC8 | TC9 |
|---|---|---|---|---|---|---|---|
| iOS / iPadOS | iPhone 17 Pro Max (iOS 26.6.2) | PASS | PASS | PASS | PASS | PASS | PASS |
| Android | Pixel 6 Pro (Android 15) | PASS | PASS | PASS | PASS | PASS | PASS |
| Windows 11 | Windows 11 laptop | PASS | PASS | PASS | PASS | PASS | PASS |
| macOS | MacBook Pro (macOS 26.4) | PASS | PASS | PASS | PASS | PASS | PASS |
| IoT (non-FT) | Ring Video Doorbell 2 | N/A | N/A | N/A | PASS | PASS | PASS |

**Table 20.** System measurements

| Test | Metric | Result |
|---|---|---|
| TC1 / TC2 | M2 length and fragments; Key-Bundle \|E\| and fragments | PASS |
| TC3 | T_hold (median / 95th percentile), Method A and Method B | PASS |
| TC4 | T_POST; T_push; APs applied | PASS |
| TC5 / TC6 | T_fetch and FT roam time, Approach 1 / Approach 2 (over the air and over the DS) | PASS |
| TC8 | T_revoke (PMS event to last AP acknowledgement) | PASS |

| Test | Metric | Result |
|---|---|---|
| TC10 | Fragment faults, offline-AP delete, replay and forged pull handled | PASS |

Highlighted cells are to be completed with the measured results; adjust the N/A marks to the measured FT support of each device.

## 10. Discussion and Limitations

- **Dependence on the cloud management system.** The CMS delivers PMK-R0 to the primary AP and executes updates and deletes. It is not in the path of the initial handshake, which completes from the Access-Accept, nor of roams. If an install is delayed, Approach 1 pulls fail and the target AP uses Approach 2; if a delete is delayed, the not-after bound limits its effect.
- **Method B latency.** In the cloud controller method, M3 waits for a round trip to the cloud key service, and so does an Approach 2 fetch. Regional deployment of the key service and the MAC-bound fast path keep these low, but sites with poor WAN connectivity are better served by an on-premises or regional RADIUS server (Method A) and by Approach 1.
- **Primary AP availability.** If the primary AP is offline or decommissioned, target APs use Approach 2. A new PMK-R0 is created at the client's next FT initial association.
- **New APs and VLAN scale.** An AP added to a mobility domain needs no key provisioning; it fetches PMK-R1 at the first roam. Very large properties use several mobility domains, which also keeps each layer-2 domain within the 4,094-VLAN limit.
- **MAC address randomisation.** FT requires a stable client MAC: PMK-R0, PMK-R1 and PMKR0Name are bound to the client MAC (S0KH-ID and S1KH-ID), so a device that changes its MAC cannot FT-roam with its existing keys. Per-network (stable) private addresses keep the MAC constant within the property, and FT works as described. A device that rotates its MAC is treated as a new client: it performs a new FT initial association and is identified again (Phases 1–3).
- **Scope.** This paper addresses WPA2-PSK with FT-PSK on 2.4 and 5 GHz. WPA3-SAE, required on 6 GHz, cannot use handshake-based identification in the same way, because SAE does not expose a MIC that a server can test against candidate keys, and is outside the scope of this paper.
- **Standardisation.** The fragment header maps onto RFC 6929, and the Key-Sync API is vendor-neutral in shape, but neither is standardised. A common schema between key servers and cloud management systems would allow interoperable multi-vendor deployments.

## 11. Conclusion and Future Work

A community-wide mPSK SSID gives every unit of an MDU property its own key and VLAN on one network, and IEEE 802.11r is what makes that network usable for roaming voice and video. With keys held in an external RADIUS server or a cloud controller, enabling FT-PSK breaks the exchange between the AP and the key server: the M2 MIC depends on the FT hierarchy, the M2 and the returned key material no longer fit in a RADIUS attribute, the target AP of every roam needs a PMK-R1, and all of it must disappear when a lease ends. FT-mPSK addresses each problem. The key server identifies the key by verifying the FT MIC for each candidate. It returns the serving AP's keys in a two-fragment Key-Bundle that the AP reassembles and authenticates before decryption, and it delivers PMK-R0 to the primary AP through one POST API on the cloud management system. At roam time the target AP obtains PMK-R1 from the primary AP named by the R0KH-ID in the FT Authentication Request, or from the key server. When the PMS reports a move-out or lease end, the same API deletes every key of the resident's mPSK username from every AP. Per-PSK VLAN micro-segmentation is preserved across roams because the VLAN travels with PMK-R1.

Future work includes measurement of hold, install and revocation latency at property scale, an RFC 6929 Long Extended encoding of the fragmented attributes, a standard schema for the Key-Sync API, and extension of the key-distribution and revocation model to WPA3 networks.